\newif\ifShowKeys
\ShowKeystrue

\documentclass[11pt,a4paper]{article} 
\pdfoutput=1
\usepackage[no-natbib-sort]{my-jheppub}

\usepackage{amsmath, amssymb}

\usepackage{mathpazo}
\usepackage{bm}
\usepackage{environ}
\usepackage{mathrsfs}
\usepackage{array,arydshln}
\usepackage{booktabs}

\usepackage{graphicx,epsfig}
\usepackage{epic}

\usepackage{tensor} 							

\usepackage{float}
\usepackage[dvipsnames]{xcolor}
\definecolor{maroon}{rgb}{0.8,0.3,0.}

\usepackage{slashed}
\usepackage[nodayofweek]{date time}

\usepackage{hyperref}

\usepackage{aurical}
\usepackage[T1]{fontenc}

\usepackage{framed}
\definecolor{shadecolor}{RGB}{255, 230, 204}

\allowdisplaybreaks

\newmuskip\pFqmuskip

\newcommand*\pFq[6][8]{%
  \begingroup 
  \pFqmuskip=#1mu\relax
  \mathcode`\,=\string"8000
  \begingroup\lccode`\~=`\,
  \lowercase{\endgroup\let~}\pFqcomma
  {}_{#2}F_{#3}{\left[\genfrac..{0pt}{}{#4}{#5};#6\right]}%
  \endgroup
}

\newcommand*\pFtildeq[6][8]{%
  \begingroup 
  \pFqmuskip=#1mu\relax
  \mathcode`\,=\string"8000
  \begingroup\lccode`\~=`\,
  \lowercase{\endgroup\let~}\pFqcomma
  {}_{#2}\widetilde{F}_{#3}{\left[\genfrac..{0pt}{}{#4}{#5};#6\right]}%
  \endgroup
}

\newcommand{\pFqcomma}{\mskip\pFqmuskip}

\newcommand{\be}{\begin{equation}}
\newcommand{\ee}{\end{equation}}

\newcommand{\mc}{\mathcal }

\newcommand{\bv}{\begin{verbatim}}
\newcommand{\ev}{\end{verbatim}}

\newcommand{\la}{\label}

\newcommand{\eps}{\epsilon}

\newcommand{\E}{\text{E}}

\def \g  {\gamma}\def \te {\textstyle} 

 \def \wz  {\widehat\zeta}\def \z   {\zeta} 

\def \ov {\over}
\def \ci {\cite}
\def \foot {\footnote}

\def \b{\beta}
\def \m {\mu}
\def \n {\nu}
\def \del{\partial}

\def \ep{\epsilon}
\newcommand{\rf}[1]{(\ref{#1})}

\def \iffa {\iffalse}
\def \d {\partial} 

\def \a  {\alpha}
\def \aa {{\rm a}}
\def \cc {{\rm c}} 
\def \ha {{1\ov 2}}

\def \g  {\gamma} \def \HH {{\mathbb H}}

\newcommand{\renyi}{{R\'enyi} }

\title{$C_T$ for 
  conformal higher spin fields   from partition function on 
  conically deformed  sphere}

\author[a]{Matteo Beccaria} 
\author[b]{ and \ \ Arkady A. Tseytlin\footnote{Also at Lebedev Institute, Moscow.}}

\abstract{
\ \ We consider  the    one-parameter   generalization $S^4_q$  of  4-sphere    with  a 
conical singularity due to identification $\tau=\tau + 2 \pi q$  in one  isometric angle.
We compute the value of the spectral  zeta-function at zero $\wz(q) = \zeta(0; q)$
that controls  the coefficient of the logarithmic UV divergence  of the  one-loop 
partition function 
 on $S^4_q$.  While the  value  of  the conformal anomaly a-coefficient   is proportional to $\wz(1)$, 
we  argue  that in general the  second  $\cc \sim C_T$   anomaly  coefficient  is related 
to  a  particular combination of the second and first derivatives of $\wz(q) $ at $q=1$.  The universality 
of this relation for $C_T$  is supported also by examples in 6 and 2 dimensions.  We use it to compute the c-coefficient 
for conformal  higher spins  finding that  it coincides  with the  "$r=-1$"   value of the  one-parameter Ansatz  suggested  in \href{https://arxiv.org/abs/1309.0785}{arXiv:1309.0785}.  
Like  the sums  of $\aa_s$ and $\cc_s$  coefficients,   the regularized sum of  $\wz_s(q)$ 
over  the whole  tower of conformal higher spins $s=1,2, ...$   is found to vanish, implying 
   UV finiteness   on $S^4_q$    and thus also  the vanishing of the  associated   R\'enyi  entropy. 
 Similar conclusions are found to apply  to  the standard 2-derivative massless  higher spin tower. 
We also   present an independent computation of the full set of  conformal anomaly  coefficients 
of the 6d Weyl graviton theory defined by a particular combination  of the three  6d Weyl invariants 
that has a $(2,0)$ supersymmetric extension. 
}

\affiliation[a]{Dipartimento di Matematica e Fisica Ennio De Giorgi,\\
Universit\`a del Salento \& INFN, Via Arnesano, 73100 Lecce, 
Italy} 

\affiliation[b]{The Blackett Laboratory, Imperial College, London SW7 2AZ, U.K.}
                       
\emailAdd{matteo.beccaria@le.infn.it} \emailAdd{tseytlin@imperial.ac.uk}

\begin{document}


\begin{flushright}\small{Imperial-TP-AT-2017-{07}}\end{flushright}				

\maketitle
\flushbottom

\section{Introduction}

In this paper we revisit the  question about  conformal anomaly  c-coefficients for 
conformal higher spin  (CHS)  fields   previously addressed in \ci{Tseytlin:2013jya,Beccaria:2017nco}.
Conformal higher spins in 4d   \cite{Fradkin:1985am,Tseytlin:2002gz,Segal:2002gd,Bekaert:2010ky,Beccaria:2014jxa,Beccaria:2016syk}
 have higher  derivative kinetic terms $h_s \del^{2s}h_s $ ($h_s$ are totally symmetric rank $s$ tensors) 
 and thus   their  generalization to a curved  metric background
 is a non-trivial  question (cf. \ci{Tseytlin:2013jya,Nutma:2014pua,Grigoriev:2016bzl,Beccaria:2017nco}).  
 A  curved  space generalization is required
 in order to  compute the corresponding conformal anomaly   coefficients   appearing  in the   (one-loop)   effective   action 
 \begin{align}
\la{1.1} 
& \Gamma = -\log Z = 
-  B_{4}\,  \log \Lambda_{_{\rm UV}} \ \  + \ \ \text{finite} \ , \\
 \la{1.2}
& B_4= \int d^{4}x\,\sqrt{g}  \ b_4 (x),\qquad  \qquad  \overline b_{4} = {(4\pi)^{2}}  b_4=
 -\text{a}\,R^{*}\,R^{*}+\text{c}\,C^{2}  \ . 
\end{align}
Here $R^{*}\,R^{*}$ is $32\pi^{2}$ times the Euler density  and $C^2$  is the square of the Weyl tensor. 
 To compute the a$_s$-coefficient   for spin $s$ field  it is enough to know the corresponding  Weyl-covariant 
 $\nabla^{2s}+...$ operator 
 on a 4-sphere  where it takes a simple factorized  form of a product of $s$  ``partially-massless"  2nd order Laplacians 
 \cite{Tseytlin:2013jya,Metsaev:2007rw,Metsaev:2014iwa,Nutma:2014pua}. 
 As a result, one finds  \cite{Giombi:2013yva,Tseytlin:2013jya} 
\begin{align}
\la{1.3}
\text{a}_{s} = \tfrac{1}{720}\,\nu\,\big(14\,\nu^{2} + 3 \nu \big), \qquad \qquad \nu\equiv s(s+1) \ , 
\end{align}
where $\nu$ is   the number of dynamical degrees of  freedom of a  spin $s$  CHS field. 
Remarkably, the  total a-anomaly defined as the  finite   part of the regularized   sum $\sum_{s=1}^\infty   e^{-\eps\,  s} \aa_s$  vanishes 
\cite{Giombi:2013yva,Tseytlin:2013jya}.\foot{This vanishing holds also in the more natural regularization 
 $\sum_{s=1}^\infty   e^{-\eps\,  (s + {1\ov 2})} \aa_s$  consistent with AdS/CFT  \ci{Giombi:2014iua}. 
Besides, the regularized  finite part of the total  CHS  partition function on $S^4$ 
 is also trivial, i.e.    \cite{Beccaria:2015vaa}, i.e. $\log Z_{\rm CHS}=\sum^\infty_{s=1} \log  Z_s =0$.}
 The factorization of the Weyl-covariant   CHS  kinetic operators   applies  for  any  conformally flat background, e.g., 
 on $S^1_q  \times S^3$,   
 where  at large $q=2\pi \beta$ one finds  that  the corresponding free energy 
   proportional to the Casimir energy  $E_c$  on $\mathbb R\times S^3$ is given by  
   \ci{Beccaria:2014jxa}~\foot{One
   finds  again that the finite part of  $\sum_{s=1}^\infty   e^{-\eps\,  (s + {1\ov 2})}  E_{c,s} $  vanishes.}
 \begin{align}
\la{1.4} 
 \ -\log Z (S^1_q  \times S^3)\Big|_{q\to \infty}   =    2 \pi q \,  E_{c} + ... \ , \qquad \qquad 
  E_{c,s} = \tfrac{1}{720}\,\nu\,\big(18 \nu^2  - 14 \nu -11\big)  \ . 
\end{align}
To determine the value of the  $\cc$-coefficient in \rf{1.2}  (which is  proportional to the coefficient $C_T$ in the 
2-point function of the flat-space stress tensor)   
  one is to consider a more  general non conformally flat   background.
  Assuming 
  that $\cc_s$ has a 
  similar  cubic $\nu$-polynomial   structure as $\aa_s$ in \rf{1.3} and   reproduce  the known values 
  for $s=1$ \ci{Duff:1977ay} and $s=2$ \ci{Fradkin:1981iu,Fradkin:1985am}, one is led to the following Ansatz 
 \cite{Tseytlin:2013jya}
\be
\la{1.5}
{\rm c}_{s} = \tfrac{1}{1080}\nu \big[43\nu^2 -59\nu+\ r\,(\nu-2)(\nu-6)\big]\ , 
\ee
where $r$ is a free parameter. 
 If one  further assumes  that  all $s>2$ 
CHS  kinetic operators factorize  into 2nd-order Lichnerowitz-type operators 
also on a Ricci flat background  (like  it  happens for the $s=2$ Weyl graviton) 
 one then  finds the expression \rf{1.5}  with $r= {1\ov 2}$  \cite{Tseytlin:2013jya}. 
 
However, the assumption  of factorization on a generic Ricci-flat   background is expected to fail in general \ci{Nutma:2014pua}.
Moreover,   when the  Weyl tensor   is non-zero,  different spins appear to mix in the kinetic term 
\ci{Grigoriev:2016bzl}  and the mixing terms lead to additional contributions to the total c-anomaly \ci{Beccaria:2017nco}. 
Also,  the  sum of $\cc_s$ with $r=\ha$  in \rf{1.5} regularized   with $e^{-\ep\, s}$  does not vanish;  
it vanishes  if  instead one chooses $r=-1$  value  \cite{Tseytlin:2013jya}.\foot{In fact, 
the finite part of $\sum_{s=1}^\infty \cc_s\,  e^{-\ep\, (s + {1\ov 2} )}$  vanishes for any value of $r$. However, 
 the special  value $r=-1$ is still selected   by demanding the consistency 
in the   results for the total $\aa$ and total $\cc$  
in the ``minimal"  case of a tower of even higher  spins  only  where they  should 
be opposite to   the values for a complex 4d scalar   for consistency with   what happens for massless
higher spins  in AdS$_5$   
(see footnote 9 in  \ci{Giombi:2014iua}).}
The expression \rf{1.5}   with $r=-1$ was also  shown to be selected   by the 
consistency with the AdS/CFT~-~related correspondence between 
massless higher spin  partition functions  in (asymptotically) AdS$_5$  space and the 
 conformal higher spin partition functions at the 4d boundary 
\cite{Beccaria:2014xda}. 

Below   we will provide a strong  independent evidence that  the $r=-1$ value of  \rf{1.5} 
\be
\la{1.6} 
{\rm c}_{s} = \tfrac{1}{360}\nu \big(14\nu^2 -17\nu-4\big)\ ,
\ee
is indeed the correct value of the c-coefficient  for  the conformal higher spin fields. 
The main idea will be  to extract the value  of $\cc_s$ from the   CHS  partition function 
on  a 1-parameter deformation  of  the 4-sphere $S^4_q$   which is an Einstein space   with a conical singularity 
\begin{align}
\la{1.7}
& ds^{2}_q = d\theta^{2}+\cos^{2}\theta\,d\tau^{2}+\sin^{2}\theta\,  d\Omega^2 \ , \qquad \qquad d\Omega^2= d\alpha^{2}_1+\sin^{2}\alpha_1 \,d\a_{2}^{2} \ , 
 \\
& \theta\in[0,\tfrac{\pi}{2}],\ \ \ \ \   \ \tau\in[0,2\pi\,q],\qquad  \qquad 
\qquad \qquad \   \alpha_{1}\in[0,\pi], \ \alpha_{2}\in[0,2\pi] \ .  \notag
\end{align}
Here the deficit angle $2\,\pi\,(1-q)$ of the $\tau$  coordinate  implies the presence  of 
a conical singularity on  a $S^{2}$ submanifold. For $q$=integer   this is a multiple cover of a sphere  while for $\g\equiv q^{-1}$=integer this 
may be interpreted as  an orbifold $S^4/\mathbb Z_{\g}$. 

The key observation is that since  $S^4_q$  is locally conformally flat (away from conical singularity), one may 
assume   that  the  CHS   kinetic operator   defined on $S^4_q$ 
still factorizes as it does on $S^4$, so that the  expression for the partition   function in terms of the 
contributions of determinants of 2nd order operators  is then 
  "inherited"  from the $S^4$ case.\foot{Alternatively, one may  define  the corresponding heat kernel 
in terms of the one  on $S^4$   using Sommerfeld-type    "orbifold" or "sum over images"  construction.} 
At the same time, having the Weyl tensor   being non-zero at the singular subspace  should allow one 
 to extract the  value of the c-coefficient    from  the $q$ dependence of the  $B_4$ coefficient of the UV divergence in \rf{1.1}.


Note that the expression for $B_4$ in  \rf{1.2} applies to regular   geometries  while in the presence of  conical singularities 
 there will be  additional "surface" terms    \ci{Fursaev:1994in,Fursaev:1995ef,Apps:1997zr}   with a   
 non-trivial dependence on $q$   entering  effectively through boundary conditions. As a result, the coefficient $B_4$ 
in \rf{1.2}  will become a non-trivial function of $q$.  
 The  log UV divergent contribution of  the  determinant  of a  single 2nd order Laplace-type operator 
 can be  computed as  the value of the corresponding 
spectral zeta function at zero, i.e.\foot{Note that $B_4$  as coefficient of the logarithmic divergence  receives contribution 
from all (zero and non-zero) modes on the Laplacian,  so that $\wz$   may  need to be corrected by the   contribution of the zero modes; here we formally assume that this correction is already taken into account.
\la{foot:zero1}} 
 \be \la{1.8} 
 B_4(q) = \wz (q) \ , \ \ \qquad  \ \ \ \   \wz(q) \equiv \z(0; q) \ , \qquad \qquad \qquad 
 \zeta(z; q) = \sum_{i}    {\rm d}_i \, \big[ \lambda_{i}(q)\big]^{-z} \ , \ee
 where ${\rm d}_i$ are degeneracies of the eigenvalues $\lambda_i$.
The general structure of  $\wz(q)$   for a bosonic field  will be as follows
\be \la{1.9} 
\wz(q) = - \,\frac{\nu}{360\, q^{3}}+\frac{p_2}{q^{2}}
+\frac{p_1}{q}+    p_0  - 2 E_c \,q\  ,
\ee
where $\nu$ is the number  of physical degrees of freedom (equal to  1 for a real $\del^2$ scalar and 
$s(s+1)$ in the  bosonic CHS  field as in \rf{1.3})
and $E_c$ is the corresponding Casimir energy on $S^3$. 

Assuming one is  able to compute $\wz(q)$  it  remains  to extract the conformal  anomaly  a- and c- coefficients from it. 
Since  the  $q=1$  case  corresponds to the regular sphere 
$S^4$  when $B_4= -2 \aa \chi(S^4) = - 4 \aa$  we  should have\foot{This  relation  is true  for the final  expression for $\wz$   taking into account possible zero modes 
 arising from  decomposition of fields  into  transverse and longitudinal parts, see   footnote \ref{foot:zero2}.} 
\be
\la{1.10}
\text{a} = -\tfrac{1}{4}\,\widehat\zeta(1) \  .
\ee
We  shall  propose  that  the expression for c in terms of $\wz(q)$  should read 
\be
\la{1.11}
\text{c} =   -\tfrac{1}{4}\,\big[q\, \widehat\zeta(q)\big]''\Big|_{q=1}
=    -\tfrac{1}{4}\, {\wz}''(1)   -   \tfrac{1}{2}  { \wz}'(1)  \ , 
\ee
where ${ \wz}' (q)\equiv  {d\ov d q } \wz(q)$.
As  we shall see,  the same  relation ($k_d$ is dimension-dependent  normalization factor) 
 \be 
 \la{ccc}
 C_T = k_d  \big[   q\,  \wz(q)\big]''\Big|_{q=1}  \ , \ee
 is  true also  in  $d=2$  \rf{ap5}    and  $d=6$ \rf{63}  cases 
 where $\cc$   is replaced by the corresponding $C_T$  coefficient 
 (proportional to  $\aa= {1\ov 3} c $ in 2d   and $\cc_3$ in 6d).  This   suggests 
  its universal  validity. 

The relation  \rf{1.11}   can be directly  verified in  all  low-spin ($s\leq 1$)   cases.  
Since $S^4_q$ is conformally related  to $S^1_q \times \mathbb H^3$ space, 
in the low-spin cases where 
 $\wz'(1) =0$,    eq. \rf{1.11}  becomes   equivalent  the 
relation derived in \ci{Perlmutter:2013gua}.
As it turns out,  for  higher-spin $s\geq 2$  cases  $\wz'(1) \not=0$      and  thus 
it is  the relation \rf{1.11} that should be applied.\foot{One feature
that distinguishes  the cases   with $s\leq 1 $  from $s\geq 2$  conformal  fields  is that 
according to Appendix  D of \cite{Beccaria:2017nco}
 for  $s\geq 2$  CHS field in flat space it is necessary 
to use the equations of motion to prove gauge invariance of the improved
symmetric traceless stress tensor.  This may  be related to a non zero value
of  the one-point function $\langle T_{\m\n}\rangle$  on $S^4_q$   or to  $\widehat\zeta'(1)\neq 0$, 
suggesting a modification of the argument in \ci{Perlmutter:2013gua}  for  $s\geq 2 $. 
} 

Intuitively, the reason   why  $\cc$  should be related to ${\wz}''(1)$   can be understood from the fact that 
$\cc\sim C_T$   should be  proportional  to the 2-point function of the stress tensor  which itself should  be given by the second variation of the effective 
action over the metric, i.e. the second term in the  expansion in the small  deformation  of  the metric \rf{1.7} 
away from the sphere  $q=1$  case.

One  may
 be tempted to represent $B_{4}(q)$ as in (\ref{1.2}), {\em i.e.} as a curvature integral
with the a and  c as  coefficients of different geometrical invariants.
 However,  $S^{4}_{q}$ has a singular curvature on $S^{2}$, and at best one  may   hope to get 
$
B_{4}(q) 
= q\,\int_{S^{4}\backslash S^{2}}b^{\rm bulk}_{4}+\int_{S^{2}}b_{4}^{\rm surf.}(q),
$
where $b^{\rm bulk}_{4}$ is as in (\ref{1.2}) and is evaluated on a smooth metric, while $b_{4}^{\rm surf.}(q)$
   non-trivially  depends  on  $q$ and   invariants of $S^{2}$.
Ref.  \cite{Fursaev:1994in}    gave  an explicit analysis of the conformal scalar operator on a  singular manifold  
 $
\mc M_{q} =  \mathscr C_{q}\times \Sigma$,
where $\mathscr C_{q}$ is a  flat cone with metric $ds^{2} = dr^{2}+r^{2}d\varphi^{2}$, 
$0\le \varphi<2\pi\,q$, while $\Sigma$ does not depend on 
 $\varphi$. 
The   above splitting of $B_{4}(q) $   into "bulk" and "surface" parts can  then  be proved  and also checked to be 
in agreement with the  expression of  $B_{4}(q)$  in terms of  the     spectral  zeta-function
  \cite{Apps:1997zr} (see also  Appendix \ref{BBB} below). It is important to stress that  in general the 
 surface term $b^{\rm surf.}(q)$ has coefficients depending on $q$ in a non-universal way,
{\em i.e.} its dependence on the spin of the field is only partly encoded in the values of the a and c
coefficients.  A major simplification occurs
at first order in  expansion in small $1-q$ and for low spin $s=0,\ha,1$. In these cases
it is possible to use the integral density in (\ref{1.2}) and take the singular
manifold into account by a delta-function contribution to the curvature 
\cite{Fursaev:1994ea,Fursaev:1995ef} (see also \cite{Solodukhin:2008dh,
Astaneh:2014sma,Fursaev:2013fta}). However, this is not true at higher orders in $1-q$ 
(and even  at leading order for bosons with spin $s\ge 2$ \cite{Fursaev:1996uz})
and this  seems to   prevent one  from obtaining the    general expression for  the  c-coefficient 
 in this approach 
 in a straightforward way.\foot{ 
 For completeness, let us mention  that if the expression of $\widehat\zeta(q)$ for a certain field
 were available as a function of the space dimension, then (\ref{1.11}) could be cross checked 
 against the representation of the 
 surface contribution as a linear combination of specific conformal invariants on $\Sigma$, 
 see for instance \cite{DeNardo:1996kp}. }

\

The rest of this paper is organized   as follows. 
In section 2  we shall  find   $\wz(q)$  \rf{1.8}  for the Laplace-type spin $s$ operators   on $S^4_q$ that enter
the partition function  of CHS fields.
 We shall  first explicitly  determine the eigenvalues  and  their degeneracies 
for $s=0,1,2,3,4$ cases (with generalization to $s\geq 5$ discussed in Appendix \ref{app:den}). 
We shall then compute $\wz(q)$  following the method discussed in Appendix \ref{app:mass}. 
The total expressions  for the $s=1,2,3,4,5$  CHS fields are presented in \rf{201}--\rf{2.15},\rf{a6}.
In subsection 2.3 we shall   discuss the   general   structure \rf{1.9} of $\wz(q)$ 
 relating it to the free energy on $S^1_q \times \HH^3$.  
 
In section 3   we shall  use $\wz(q)$ to determine the conformal   anomaly coefficients  corresponding to the CHS fields. 
We shall  discuss  the relations \rf{1.10} and \rf{1.11} and comment on similar relations following from \renyi entropy. 
We shall also determine  the general expression for $\wz_s(q)$   for any value of  CHS spin $s$   satisfying non-trivial 
consistency  conditions. Like for the sums  of $\aa_s$ and $\cc_s$  coefficients  the regularized sum of  $\wz_s(q)$ 
over  the whole  tower of conformal higher spins  is found to vanish, implying 
that the full CHS theory is one-loop UV finite  on $S^4_q$ space   and thus  
implying as well the vanishing of the total \renyi entropy. 
 Similar conclusions apply  to  the standard 2-derivative massless  higher spin tower  (see Appendix \ref{JJ}).

In section 4 we shall   discuss  generalization to 6 dimensions. We shall  compute 
the corresponding $\wz(q)$  for low-spin $s=0,1,2$ CHS fields  on $S^6_q$    space  and  show that the  expected relation \rf{63} for the $\cc_3 \sim C_T$   conformal anomaly coefficient  is   fully consistent not only with the  previously known    2-derivative  scalar 
 and 4-derivative vector results
   but also with the new  result for the 6d Weyl graviton  conformal anomaly computed independently
 from the Seeley-DeWitt coefficient in Appendix \ref{app:weyl}. 

The universality  of the relation for $C_T$  \rf{ccc} is  further supported by the discussion of the $d=2$ case in Appendix \ref{AA}. 
In Appendix \ref{BBB} we shall comment on the relation between  the expressions  for $\wz(q)$  for spin $s=0,1,2,3$  Laplacians   and some previous  results \ci{Fursaev:1996uz}  for the $B_2$ Seeley-DeWitt coefficient  found in  the "geometrical" approach.

\def \De  {\Delta} 
\def \ed  {\end{document}}

\section{Zeta-function of generalized  spin $s$ Laplacian on $S^4_q$} 

To compute the function $\wz(q)$ in \rf{1.8}  for  conformal higher spin  fields on $S^4_q$ 
   and thus the corresponding   a and c    anomaly coefficients  using \rf{1.10} and \rf{1.11} ,
   our  starting point  will be the  CHS   partition 
     function on $S^4$ \ci{Tseytlin:2013jya}. 
     It is  expressed in terms of the determinants  of 
     generalized Laplace  (or Lichnerowitz-type) operators on  unit-radius $S^4$    defined on a  totally symmetric transverse traceless (TT) 
     rank $s$ tensor
   \be \la{21} 
   \Delta_{s\,  \perp}(M^{2}) \equiv  (-D^{2}+M^{2})_{s\, \perp}  \ , \ee
   where $M^2$  is a constant parameter   that need not  be positive (the scalar curvature is $R=12$). 
For example, the one-loop  $S^4$ partition functions  of the standard 2-derivative 
massless conformally coupled scalar,\foot{The $s=0$   member of CHS tower in $d=4$ 
is    non-dynamical, i.e. $Z_0=1$,  but it is  useful to consider also 
the   2-derivative conformally coupled scalar  to be able to compare  with  previous results on $S^4_q$.
In what follows the $s=0$ case will stand for the  $\del^2$ scalar field.}
$s=1$  Maxwell vector, $s=2$  Weyl graviton  and $s=3$ and $s=4$   CHS   fields  read 
 \begin{align} \la{22} 
&Z_{0}=  \Big[\frac{1}{\det\Delta_{0}(2)}\Big]^{1/2} 
 \ , \qquad \qquad  Z_{1}= \Big[\frac{\det\Delta_{0}(0)}{\det\Delta_{1\perp}(3)} 
\Big]^{1/2}  \ , \\
&Z_2 =  \Big[\frac{\det\Delta_{1\perp}(-3)\,\det\Delta_{0}(-4)}
{\det\Delta_{2\perp}(4)\,\det\Delta_{2\perp}(2)}
\Big]^{1/2} \ , \la{23} \\
&Z_3 =   \Big[\frac{\det\Delta_{2\perp}(-8)\det\Delta_{1\perp}(-9)\,\det\Delta_{0}(-10)}
{\det\Delta_{3\perp}(5)\,\det\Delta_{3\perp}(3)\,\det\Delta_{3\perp}(-1)}
\Big]^{1/2}  \ ,    \la{24} \\
&Z_4 =  \Big[\frac{\det\Delta_{3\perp}(-15)\det\Delta_{2\perp}(-16)
\det\Delta_{1\perp}(-17)\,\det\Delta_{0}(-18)}
{\det\Delta_{4\perp}(6)\,\det\Delta_{4\perp}(4)\,\det\Delta_{4\perp}(0)\,\det\Delta_{4\perp}(-6)}
\Big]^{1/2} \la{25} \ . 
\end{align} 
We will assume that these partition functions  extended to   $S^4_q$   have the same product structure 
with each operator  now defined on  $S^4_q$.  Thus the problem of computing them  reduces to 
finding the  dependence of the spectrum  of the operator \rf{21} on the conical deformation parameter $q$. 

We will closely  follow  the  approach  of \cite{DeNardo:1996kp}  where the scalar and vector operators   were discussed, 
  generalizing it to  the $s>1$  case. We  will  first assume that $q=1/\gamma \leq 1 $
where  for  an  integer $\g$  the space $S^4_q$   becomes the  $\g$-quotient of $S^4$. 
For $q=\g=1$  the   spectrum  should 
reduce to  the  regular  $S^4$ one    found in \cite{rubin1984eigenvalues}. 

As explained later, for $\gamma>1$ the spectrum will in  general  be 
different in the intervals $\gamma\in [n, n+1)$. Starting from a certain $n$, depending on the spin 
$s$, the 
structure of the spectrum will be independent of $\gamma$. The relevant range for us 
will be $\gamma\in[1, 2)$ since to find the  conformal anomaly coefficients in \rf{1.10}, \rf{1.11} 
 we  will   interested in  the expansion near $q\to 1$.\footnote{If $\gamma<1$,
 there may be a finite number of normalizable eigenmodes that are, however,  singular on some 
 subspace. This has been noticed to happen already in the scalar case \cite{DeNardo:1996kp}.
 If only regular eigenmodes are
 considered, then our results extend to a neighbourhood of $\g=q^{-1}=1$.}

\def \rd {{\rm d}}

The eigenvalues $\lambda_{n,m}$ of $\Delta_{s\, \perp}(M^{2})$ in \rf{21} 
will  be  parametrized by the  two integers $n,m\ge 0$  as 
\be
\la{2.1}
\lambda_{n,m}(\gamma) = (n+\gamma\,m)(n+\gamma\,m+3)-s+M^{2} \ . 
\ee
The 
 degeneracies ${\rm d}^{(s)}_{n,m}$ 
may be found using  the correspondence between $\De_{s\, \perp}(M^{2})$
and the Laplacian  on  the ambient flat space 
with coordinates $(x^{1},x^{2},x^{3}; x^{4},x^{5})$ and the constraint $|{x}|^{2}\equiv x^a x^a =1$
with the conical   singularity implemented by the   identification $x^a(\tau) = x^a(\tau + 2 \pi q)$ 
where $\tau$ is the coordinate in \rf{1.7}
(see \cite{DeNardo:1996kp}).

\iffa 

logic is discussed in the paragraph after 2.2 of https://arxiv.org/pdf/hep-th/9610011.pdf
I am not sure this is the only way to construct eigenfunctions, but at least for coding it is very convenient to 
work in the embedding R^d+1 space. Anyway your question about cone and embedding space is 
definitely non trivial. In the paragraph mentioned above logic is that one can consider R^d+1 with 
conical singularity by an identification x^K(tau)=x^K(tau+2pi q) and the point should be that S_q is 
embedded in R_q by same equation r^2=1

 The flat 
Laplace operator has eigenvalues related to (\ref{2.1}) by a constant shift \cite{rubin1984eigenvalues}
\be
\la{2.2}
\lambda_{n,m}(\gamma) = (n+\gamma\,m)(n+\gamma\,m+3).
\ee
\fi 
The explicit spectrum for $\g\not =1$ can be constructed by  starting   with a suitable Ansatz for the eigenstates consistent with 
periodicity on $S_{1/\g}^{4}$  generalizing to $s>1$  the   discussion of the scalar and vector cases in 
  \cite{DeNardo:1996kp}.

\subsection{Eigenvectors and  degeneracies} 

 In general, the eigenvectors of the Laplacian on  the  flat ambient space $(x^{1}, \dots, x^{5})$   will be  
  tensors $(\Phi^{a_{1} \dots a_{s}})_{n,m}$    corresponding to  the eigenvalues (cf. \rf{2.1})
\be
\la{2.2}
\hat \lambda_{n,m}(\gamma) = (n+\gamma\,m)(n+\gamma\,m+3)\ . 
\ee
They must be symmetric, traceless, and also tangential and  transverse 
in the ambient space
\be\la{289}
x^{a}\,\Phi\indices{_{a}^{i_{2} \dots i_{s}}} = 0,\qquad  \qquad \partial_{x^{a}}\,
 \Phi\indices{_{a}^{i_{2}  \dots i_{s}}} = 0.
 \ee
Here   we  split the  coordinate indices  as  $a=(i, +,-)$
 where $i=1,2,3$  and $x^{\pm} =  \frac{1}{\sqrt 2}(x^{5}\pm i\, x^{4}).$
 The ansatz  for the  tensor components    with   all indices from the 3-subspace reads 
\foot{The  constraint $|x|\equiv \sqrt{ x^a x^a}=1$  is imposed   after taking  the derivatives   when imposing the 
transversality condition and applying the Laplacian \cite{DeNardo:1996kp}.} 
\be
\la{2.4}
(\Phi^{i_{1}\dots i_{s}})_{n,m} = |{x}|^{-n+\gamma\,m}\,(x^{+})^{\gamma\,m}\sum_{p\ge 0}
(B^{i_{1}\dots i_{s}})^{m}_{i_{1}\dots  i_{s-2p}}x^{i_{1}}\cdots x^{i_{s-2p}}\,(x^{+}x^{-})^{p}\ . 
\ee
The sum over $p$ involves a finite number of terms, i.e. monomials of total degree $s$ with some
explicit power of $x^{+}x^{-}$. 
In the case of  one $\pm$ index  we have instead   
\be
\la{2.5}
(\Phi^{i_{1}\dots i_{s-1}\pm})_{n,m} = |x|^{-n+\gamma\,m}\,(x^{+})^{\gamma\,m\pm 1}\sum_{p\ge 0}
(B^{i_{1}\dots i_{s-1}\pm})^{m}_{i_{1}\dots i_{s-1+2p}}x^{i_{1}}\cdots x^{i_{s-1+2p}}\,(x^{+}x^{-})^{p},
\ee
and similar  expression is assumed  when there are  more indices of the $+$ or $-$ type.
 The  regularity of the eigentensor components with one or more
"$-$" indices, i.e. the absence of negative powers of $x^{+}$, requires the sum over $p$ to start at some positive    value 
 depending on the value of $\gamma m$.
This is the unique source of the  $\gamma$-dependence  of the spectrum.
In practice,  this is a feature that starts being relevant for $s\ge 2$.

 All eigenvectors of the form  (\ref{2.4}), (\ref{2.5})  or with more $\pm$ indices, appear
 together with a mirror copy where $x^{+}\leftrightarrow x^{-}$ when $m>0$. The solutions with $m=0$
 are automatically symmetric under this exchange.
By an explicit enumeration, we  can then determine  the degeneracies $\text{d}^{(s)}_{n,m}$.
For  the  $s=0$ and $1$   cases   we reproduce  the  results of \cite{DeNardo:1996kp}
  for  $S^d_q$   with any  $d$:

\noindent
{\bf scalar}: \ \   $ n+m \geq 0$ 
\begin{align}
\la{2.7}
&\text{d}^{(0)}_{n,m>0} = 2\,\binom{n+d-2}{d-2}\ \  \stackrel{d=4}{\to} \ \ (n+1)(n+2), \notag\\
&\text{d}^{(0)}_{n,0} = \binom{n+d-2}{d-2}\ \  \stackrel{d=4}{\to }  \ \ \tfrac{1}{2}\,(n+1)(n+2).
\end{align}
{\bf spin 1}:    \ \  $n+m \geq 1$ 
\begin{align}
\la{2.8}
&\text{d}^{(1)}_{n,m>0} = 2\,(d-1)\binom{n+d-2}{d-2}\ \  \stackrel{d=4}{\to} \ \ 3\,(n+1)(n+2), \notag \\
&\text{d}^{(1)}_{n,0} = \frac{1}{n+1}\binom{n+d-3}{d-2}\big[d^{2}+(n-4)d+5-n\big]\ \  \stackrel{d=4}{\to}\ \ 
\tfrac{1}{2}\,n\,(3n+5)\ . 
\end{align}
For  $s=2,3,4$ and $\gamma\in[1,2)$ we  find the following  results for  the 
degeneracies in $d=4$:\footnote{It is 
useful to check the correspondence with the known results 
in the regular $S^4$ limit   of $\gamma=1$. 
For general spin $s$ and $\g=1$, setting $N = n+m$ we have in 4d: \ 
$\text{d}_{N}^{(s)} = \tfrac{1}{6}\,(2s+1)\,(2N+3)(N+s+2)(N-s+1)$.  We have checked that indeed 
in all cases $\sum_{s\le n+m\le N}\ \text{d}^{(s)}_{n,m} = \text{d}_{N}^{(s)}$.
\la{foot:tot}}

\noindent
{\bf   spin 2}: \ \   $n+m \ge 2 $
\begin{align}
& \text{d}^{(2)}_{n,0} = \tfrac{1}{2}\,(n-1)\,(5n+8),\qquad
\text{d}_{n,1}^{(2)} = n(5n+11), \qquad
\text{d}^{(2)}_{n,m>1} = 5\,(n+1)(n+2).\la{28}
\end{align}
{\bf  spin 3}: \ \   $n+m \ge 3$
\begin{align}\la{281}
& \text{d}^{(3)}_{n,0} = \tfrac{1}{2}\,(n-2)\,(7n+11),\qquad
\text{d}_{n,1}^{(3)} = (n-1)(7n+16), \notag \\
& \text{d}_{n,2}^{(3)} = n(7n+17), \quad\qquad \ \ \ \ \ \ \ \ \ \  
\text{d}^{(3)}_{n,m>2} = 7\,(n+1)(n+2).
\end{align}
{\bf spin 4}: \ \  $n + m \geq 4$
\begin{align}\la{282}
& \text{d}^{(4)}_{n,0} = \tfrac{1}{2}\,(n-3)\,(9\,n+14), \qquad
\text{d}^{(4)}_{n,1} = 3\,(n-2)\,(3\,n+7), \notag \\
& \text{d}^{(4)}_{n,2} = 3\,(n-1)\,(3\,n+8), \qquad\ \ \ 
\text{d}^{(4)}_{n,3} = n\,(9\,n+23), \notag \\
& \text{d}^{(4)}_{n, m>3} = 9\,(n+1)\,(n+2).
\end{align}
We suggest a generalization  of these expressions for degeneracies to any integer $s>4$ in Appendix \ref{app:den}. 

\def \zr {\zeta_{\rm R}}

\subsection{Computation of $\widehat\zeta(q)$}

To find  the spectral zeta-function    and thus $\wz(q)$  it remains to perform the sum in \rf{1.8}.
Representing the eigenvalue \rf{2.1} for particular $s$ and $M^2$ as  
 \be \la{70}
\lambda_{n,m} = (n+\gamma\,m)(n+\gamma\,m+3)-s+M^{2} =
(n+\gamma\,m+\mu)(n+\gamma\,m+\mu')\ , \ee  
we  thus need to compute 
\be\la{71}
\sum_{n\ge 0}\sum_{m\ge 0}\text{d}_{n,m}^{(s)}\ \big[(n+\gamma\,m+\mu)(n+\gamma\,m+\mu')\big]^{-z}.
\ee
One possible approach is to  follow  \cite{Allen:1983dg}  
and expand $[\lambda_{n,m}]^{-z} $  in powers of the $n,m$ independent 
 term $-\tfrac{1}{4}(\mu-\mu')^{2}$.\footnote{Only a finite number of terms will give a non-zero contribution 
 in the limit  $z\to 0$ so  that  the final result is stable for 
a sufficiently long  expansion.\la{foot:mu}}
Doing the sum over $n$, one can then reduce the 
expression  for the  spectral  $\zeta$ function  of $\Delta_{\perp\, s}$ on $S^4_{1/\g}$  to a sum of  terms with coefficients   being the 
Hurwitz   zeta functions 
$\zeta_{\rm R}(a,b)$. Using  the integral   representation  for  $\zeta_{\rm R}(a,b)$ 
\be\la{72}
\zeta_{\rm R}(a,b) = \frac{1}{\Gamma(a)}\int_{0}^{\infty}dy\ \frac{y^{a-1}\,e^{-b\,y}}{1-e^{-y}},
\ee
one  can then  do the  sum over $m$   in the integrand,   expand in $\gamma$, integrate term by term
in $y$, and finally send $z\to 0$ to obtain 
 $\widehat \zeta(q)$  in \rf{1.8}  with $q=1/\g$.
 An alternative  more straightforward  approach (described in  Appendix \ref{app:mass}  in the $s=0$  case)  is to use the  heat  kernel representation  taking into account the  factorized form  of the eigenvalues  in \rf{71}. 
Applying this procedure,  the general expressions  for  $\widehat\zeta_{s\,\perp}(q; M^{2})$  corresponding to 
 the operator (\ref{21}) defined
on $S^{4}_{q}$  with $s=0,1,2,3,4$ 
  are    then found   to be 
\begin{align}
\widehat\zeta_{0}(q; M^{2}) &= -\frac{1}{360\,q^{3}}+\frac{2-M^{2}}{12\,q}+\frac{1}{120}\,(
19-30\,M^{2}+10\,M^{4})\,q, \notag\\
\widehat\zeta_{1\,\perp}(q; M^{2}) &= -\frac{1}{120\,q^{3}}+\frac{3-M^{2}}{4\,q}+
M^{2}-\frac{7}{3}+\frac{1}{40}\,(59-50\,M^{2}+10\,M^{4})\,q, \la{288} \\
\widehat\zeta_{2\,\perp}(q; M^{2}) &= -\frac{1}{72\,q^{3}}-\frac{2}{q^{2}}+\frac{68-5\,M^{2}}
{12\,q}+5\,M^{2}-\frac{26}{3}+\frac{1}{24}\,(119-70\,M^{2}+10\,M^{4})\,q,\notag\\
\widehat\zeta_{3\,\perp}(q; M^{2}) &=
-\frac{7}{360\,q^{3}}-\frac{14}{q^{2}}-\frac{7\,(M^{2}-53)}{12\,q}+14\,M^{2}-\frac{14}{3}
+\frac{7}{120}(199-90\,M^{2}+10\,M^{4})\,q,\notag \\
\widehat\zeta_{4\,\perp}(q; M^{2}) & = -\frac{1}{40\,q^{3}}-\frac{54}{q^{2}}-
\frac{3\,(M^{2}-150)}{4\,q}+30\,M^{2}+56+\frac{3}{40}\,(299-110\,M^{2}+10\,M^{4})\,q.\notag
\end{align}
To obtain  the total  values of $\wz(q)$  for CHS fields 
it remains to  sum up 
  the contributions from different factors in the  partition functions \rf{22}--\rf{25}.
 When combining $\wz(q)$  \rf{288}
    for the operators $\Delta_{s\, \perp}$ 
   one needs to  account for the contribution
   of  the  number $n_z$ of the artificial zero modes introduced by the splitting of the fields into 
   transverse parts, i.e.  the corrected expression is\footnote{
\la{foot:zero2}
As the original  action and thus the partition function  is expressed in terms of unconstrained fields
 one has to remove 
 spurious  zero modes related to  splitting the   rank $s$ 
 tensor  into its transverse plus longitudinal parts
(see \cite{Christensen:1979iy,Fradkin:1983mq} for the case of $s\le 2$ in $d=4$).
 This splitting  introduces  additional $n_{\rm z}$ zero modes of the Jacobian
 of the change of variables. These modes were not present in the original unconstrained operator 
 and their number must be subtracted from  $\wz_\perp $ leading to $ B_{4} =\wz=  \wz_\perp  -n_{\rm z}$.
}
\be
\la{344}
\wz(q)  =\sum_{s'}   \wz_{s'\, \perp} (q) -n_{\rm z}  \ . 
\ee
The number of zero modes associated  to   a TT spin $s$ tensor is equal to the number of 
rank $s-1$ conformal Killing tensors on $S^4$ 
 \cite{Eastwood:2002su}
 \footnote{In general  $d$,
this is the dimension of the $(s-1, s-1, 0, ..., 0)$ representation of $SO(d + 1, 1)$.}
\be
{\rm k}_{s} = \frac{1}{12}s^{2}(s+1)^{2}(2s+1) \ . \la{333}
\ee
Explicitly, for the  
partition functions  in \rf{22}--\rf{25} we get 
\begin{align}
\la{3.6}
s=0:\quad  & n_{\rm z} = 0, \qquad \qquad\qquad\qquad\qquad \ 
s=1:\quad   n_{\rm z} = \text{k}_{1} = 1, \notag \\
s=2:\quad  & n_{\rm z} = 2\text{k}_{2}-\text{k}_{1} =29 , \qquad \qquad \ \ \ 
s=3:\quad   n_{\rm z} = 3\text{k}_{3}-\text{k}_{2}-\text{k}_{1} =236, \notag \\
s=4:\quad  & n_{\rm z} = 4\text{k}_{4}-\text{k}_{3}-\text{k}_{2}-\text{k}_{1} =1100.
\end{align}
As  a result, we find  the following expressions for $\wz(q)$   for the  $\del^2$  ($M^2=2$)
  and $\del^4$  ($M^2=0,2$) 
conformal scalars  and $s=1,2,3,4$   CHS fields 
\begin{align}
\wz_\varphi (q) 
 = &
 -\frac{1}{360\,q^{3}}-\frac{1}{120}q   \ ,   \qquad    \qquad  \widehat\zeta_{\varphi^{(4)}}(q) =
-\frac{1}{180\, q^3}+\frac{1}{6\,q}+\frac{3}{20}\, q,     \la{200} \\
 \wz_1(q) =& -\frac{1}{180\,q^{3}}-\frac{1}{6\,q} -  {1\ov 3} 
 -\frac{11}{60}\,q\ , \la{201}  \\
\wz_2(q)=&   -\frac{1}{60\,q^{3}}-\frac{4}{q^{2}}+\frac{41}{6\,q} - 11 
-\frac{553}{60}\,q\ , \la{202} \\
\wz_3(q)=& -\frac{1}{30\,q^{3}}-\frac{40}{q^{2}}+\frac{227}{3\,q}  -  92  
-\frac{2413}{30}\,q\ , \la{203}  \\
   \wz_4(q)=&
-\frac{1}{18\,q^{3}}-\frac{200}{q^{2}}+\frac{1165}{3\,q}   -  \frac{1300}{3}    
-\frac{2303}{6}\,q \la{2.15} \ .\end{align}
The  standard scalar  and  spin 1  cases were discussed in \cite{Fursaev:1993hm,DeNardo:1996kp}. 
The 4-derivative scalar expression was  found in \cite{Dowker:2017qkx}.
The  vector   expression in  \rf{201}   agrees   with  the result of  \ci{DowkerNew}.
%

Similar   analysis   can be repeated in the  case of the  fermionic CHS  fields
with kinetic terms $\del^{2s}$ with $s={1\ov 2}, {3\ov 2}, ...$  \ci{Tseytlin:2013jya}.
For   $s={1\ov 2}$   fermion  with the standard $\del$-action or  conformal $\del^3$-action  
we find, using the results of \cite{Dowker:2017qkx}\foot{One is to set   $k={1\ov 2}$  and $3\ov 2$ in eqs. (18),(19)  in 
\cite{Dowker:2017qkx}. Note that $q$ in  \cite{Dowker:2017qkx}
corresponds to our $\g= q^{-1}$.   } 
\begin{align}
\la{2.17}
\widehat\zeta_{\psi}(q) = -\frac{7}{1440\,q^{3}}-\frac{1}{48\,q}-\frac{17}{480}q ,\qquad \qquad 
\widehat\zeta_{\psi^{(3)}}(q) = -\frac{7}{480\,q^{3}}+\frac{5}{48\,q}
+\frac{29}{480} q\ .
\end{align}
\iffa 
 For instance, higher derivative GJMS fields of bosonic or fermionic type
and  $2k$ derivatives in kinetic action  have \cite{Dowker:2017qkx}
\begin{align}
\text{boson}: \quad & \widehat\zeta_{\text{GJMS}, k}(q) = 
-\frac{k}{360 q^3}+\frac{k \left(k^2-1\right)}{36 q}+\frac{1}{360} k \left(6
   k^4-20 k^2+11\right) q, \quad k=1,2,3,\dots \notag \\
\text{fermion}: \quad & \widehat\zeta_{\text{GJMS}, k}(q) = -\frac{7 k}{720 q^3}+\frac{k \left(k^2-1\right)}{18 q}+\frac{1}{90} \left(-6
   k^5+20 k^3-11 k\right) q, \quad k = \tfrac{1}{2}, \tfrac{3}{2}, \dots,
\end{align}
and one can check that $\widehat\zeta'(1)=0$ for (positive) $k=1,2$
in bosonic case and $k=1/2, 3/2$ in fermionic case. These are the relevant cases in 4d. 
\fi

While  in this paper  we  are interested in  fields  defined on $S^d_q$     with even $d$, 
let us  note    that in the case of  odd $d$   one  may 
expect the  coefficient  of the  log UV divergence in \rf{1.1} 
 to   vanish (as the  space  has no boundary  
 all   log  UV   divergences   should   be  bulk ones   and  thus 
 should be built out of curvature invariants).  Thus  in odd $d$ 
one  may   expect to   find  that $\wz(q)=0$. 
Indeed, one can  check  that this is the case  for a scalar or spin 1  field using \rf{2.7} and \rf{2.8} 
(for any $M^2$   parameter in the operator and after subtracting as in \rf{344} 
the constant  $n_{\rm z}=1$ in the $s=1$ case). 
However, in the  $s=2$, $d=3$    case, i.e.  for the   operator $\Delta_{2\, \perp}(M^2) $ 
defined  on $S^3_q$  one finds 
$\wz(q) =4\,q^{-1}+6$. Subtracting 
$ n_{\rm z }= 10$  gives $\wz(q) = 4 \, q^{-1} - 4 $.
This   vanishes   as expected  for $q=1$, i.e. for a round 3-sphere, but 
is non-zero  in general. The same is then expected to  happen also  for $s>2$
and requires  an  explanation.

\subsection{General structure of $\widehat\zeta(q)$}
\la{sec:gen}
The leading  small  $q$  and   large  $q$ asymptotics of the $\wz(q)$-functions on $S^4_q$  in
 \rf{200}--\rf{2.17} have  the universal   structure \rf{1.9}, i.e. 
 \be 
 \wz(q) = -\frac{\nu }{360  \, q^3}  + ...+ ( - 2 E_c )\,  q  \  , \la{2.21} \ee
 where $\nu$ is the  number of dynamical degrees of freedom  in bosonic case (rescaled by $7\ov 8$  in the fermionic case)  and $E_c$ is the Casimir  energy on $\mathbb R \times S^3$.  Indeed, the metric \rf{1.7}  effectively simplifies
 in these limits:  for $q\to 0$ the  $\tau$-direction  shrinks to zero    (or the  transverse  3-space blows up) 
 so we get effectively $S^1_q \times \mathbb R^3$, while for $q\to \infty$ the $\tau$-direction 
   decompactifies 
 and   the  space becomes   similar to  $\mathbb R \times  S^3$. This suggests  that 
 $\log Z$   should be related to the free energy on $S^1_q \times \mathbb R^3$ for $q\to 0$
   and  on $\mathbb R \times  S^3$ for $q\to \infty$.\footnote{Let us mention in this connection 
      a discussion   \cite{Shaghoulian:2016gol} of an 
    interesting duality  between
    the $q\to 0$ and $q\to \infty$ limits  of  the partition function  on $S^1_q\times S^3/\mathbb{Z}_n$.}
   
   To make this connection more explicit  we  may use  that the   metric of $S^d_q$ is  related by 
   a conformal rescaling  (by $\cos^2 \theta$ in \rf{1.7}) to the metric of  $S^1_q \times \HH^{d-1}$ 
where $\mathbb H^{d-1}$ is a real hyperbolic space of unit curvature radius.\footnote{This conformally mapping   has an important role
in the discussions of  \renyi entropy, see,  e.g., \cite{Klebanov:2011uf}. }  
The effective actions  on $S^d_q$  and on $S^1_q \times \HH^{d-1}$ are then related by a  finite 
integrated conformal anomaly term. This 
 allows one  to relate $\widehat\zeta(q)$ on $S^{d}_{q}$ to the 
thermal free  energy  of a  CFT on $S^{1}_{q}\times \mathbb H^{d-1}$ where 
 the length of the  thermal circle    is $\beta=2\pi q$.

\iffa 
The relation between $S^{4}_{q}$ and $S^{1}_{q}\times S^{3}$
is particularly interesting because for a CFT the relevant parameter is the ratio $q$ of the radii of the
thermal cycle $S^{1}_{q}$ and the spatial sphere $S^{3}$. In particular, the limits
$q\gg 1$ should be related to the spectrum on $\mathbb R \times S^{3}$, while 
and $q\ll 1$ should capture thermal behaviour in flat space, {\em i.e.} on $S^{1}_{q}\times \mathbb R^{3}$.
\fi

In  the case of   the   homogeneous   space    $S^{1}_{q}\times \mathbb H^{3}$   
the free  energy $F(q) $  is proportional to its volume
$ 2\pi q\,  \text{Vol}({\mathbb H^{3})} $. Extracting the IR divergent   factor in the volume, we may 
define the  IR finite  "free energy" 
$\mc F(q)$  by 
\be
\la{2.18}
F(q)  \equiv  \mc F(q) \,  \log \Lambda_{\rm IR} \ . 
\ee
\iffa 
For even $d$  the free   energy   on  ${S^{1}_q\times \mathbb H^{d-1}}$   
 does not contain logarithmic UV   divergences\foot{
Since 
$S^{1}_{q}$ factor is flat and $\mathbb H^{d-1}$  is conformally flat,   all 
logarithmic   divergent terms  containing the Weyl tensor vanish, while the   Euler density in $d$  dimensions  vanishes 
when evaluated on $\mathbb H^{d-1}$.}
 \fi 
   Recalling that  $\wz(q)$  is the coefficient of the log of the UV cutoff  (cf. \rf{1.1},\rf{1.8}), 
 restoring the dependence on the curvature radius $\rm r $  and comparing the  coefficients 
 of $\log \rm r$    suggests a  direct relation between $\wz(q)$  and $\mc F(q)$,  or   explicitly
 $
\widehat\zeta(q) = -\mc F(q)  .  
$
 For $q\to \infty $  the  free energy of  $S^1_q \times \HH^3$   should approach the one  on
  $\mathbb R \times \HH^3$.\foot{Note that 
 in a conformal theory the partition function depends on the ratio of the scales of $S^1$ and $\HH^3$.}
 Since $\HH^3$ is related by the  analytic continuation  to $S^3$, that implies  that 
 $\mc F(q\gg 1) \to  2 E_c  q $
 where $E_c$ is the Casimir energy on $S^3$.\foot{The proportionality coefficient  can be understood as 
 follows  \cite{Dowker:2017qkx}:  as  $\text{Vol}(\mathbb H^{3})=-2\pi\log \Lambda_{\rm IR}$ and
$\text{Vol}(S^{3}) = 2\pi^{2}$,     there is a relative  $-  {1\ov \pi} $  factor that 
free energy on $S^1_q \times S^3$  in \rf{1.4} should be multiplied by. }
 
 
In general, $F(q)$ computed on $S^1_q \times S^3$   or $S^1_q \times \HH^3$
contains a non-universal  (UV power-divergent)  part   proportional to the volume  and thus  linear in $q$
and a  universal finite part.  One  may 
 define $F(q)$  in a particular scheme   where  all    non-universal 
 power UV divergences  
 are   subtracted out 
 and the linear in $q$ part is  the Casimir energy, i.e. 
   \be
\la{2.19}
\widehat\zeta(q) = - {\mc F}(q)= -\overline {\mc F}(q)-2\,E_{c}\,q \ , \ \ \ \ \ \   
\ee
where $\overline {\mc F}(q)$  contains only non-positive powers of $q$.
The function 
 $\overline {\mc F}(q)$  was  computed
 for free  conformal fields  with spins $s\le 1$,  including higher derivative cases,  in  
  \cite{Klebanov:2011uf,Beccaria:2017dmw}.\foot{ The relation \rf{2.19}  is valid also for a generic 
GJMS conformal higher derivative scalars  \cite{Dowker:2017qkx}.
Note that $\overline {\mc F}(q)$  was not so far  computed directly on $S^1 \times \HH^3$ for $s \geq 2$:  it is non trivial to extend the analysis of \cite{Beccaria:2017dmw} to  spins  higher than 1 due to several ambiguities  discussed there. 
}

In general, the definition of 
$E_c$ on $S^3$   is  scheme-dependent  -- it depends  on the definition of the stress tensor 
or the coefficient  $g$ of the total derivative $D^2R$   term in $\langle T^m_m \rangle$, 
i.e. $E_c = { 3 \ov 4} (\aa  +   {1\ov 2}  g)$   \ci{Cappelli:1988vw}. 
A natural scheme  is  the one when  $E_c$   is   determined
from the  single particle partition function  of the corresponding CFT   using  the standard  zeta-function 
definition (see, e.g.,   \ci{Beccaria:2014jxa,Beccaria:2014xda}). 
It is this  $E_c$  that   appears  as the $q$-coefficient  in $\wz(q)$  \rf{1.9},\rf{2.21}
($\wz(q)$  itself is  scheme-independent  being the coefficient of log UV divergence on $S^4_q$).


\iffa
Thus  while $\wz(q)$   should be scheme-independent, the separation of the two terms in \rf{2.19}  may be ambiguous.
At the same  time,  in  the $-2E_c q$ term in $\wz(q)$ in \rf{2.21} 
$E_c$   is  assumed  to be  determined
from the  single particle partition function  of the corresponding CFT   using  the standard  zeta-function 
definition  and in that sense is  defined unambiguously.
\label{foot:Ec}
just to summarize again:
say F on   Rx S3 or  Rx H3   has q-linear
non-universal part plus   universal rest;
dropping q-term is a particular scheme.
that  agrees  with E_c  as coeff  of q as  being scheme dependent.
at the same time we should say that
in zeta(q)    q-coeff (call it A)
 is such that A= - 2E_c   where E_c is  Casimir en on S3 in special scheme
-- obtained from one-particle part funct.
\fi

  
 One can  indeed  check that   the order $q$  coefficients in \rf{200}--\rf{2.17}  are the corresponding values of $E_c$  for the  conformal  $\del^2$  and $\del^4$  scalars,  $\del$   and $\del^3$ fermions 
 and $s=1,2,3,4$ CHS  fields 
 summarized   below (see also \rf{1.4})
\be\la{2.29}
\renewcommand{\arraystretch}{1.5}
\begin{array}{ccccccccc}
\toprule
 & \varphi & \varphi^{(4)} & \psi & \psi^{(3)} & \text{CHS}_{1} & \text{CHS}_{2} & \text{CHS}_{3}
 & \text{CHS}_{4} \\
 \midrule
 E_{c} \phantom{\quad}
  & \frac{1}{240} & -\frac{3}{40} & \frac{17}{960} & -\frac{29}{960} & \frac{11}{120} & \frac{553}{120}
 & \frac{2413}{60} & \frac{2303}{12} \\
\bottomrule
\end{array}
\ee

In the opposite  $q\to 0$  limit  the  free energy on  $S^1_q \times \HH^3$   should approach the one 
 on $S^1_q \times \mathbb R^3$, i.e.   should have the same  $q\to 0$ asymptotics  as the 
 thermal free energy on $S^{1}_{q}\times S^{3}$  (see, e.g.,  \cite{Kutasov:2000td}). 
 Thus it should   simply be proportional to the free energy of 
 a single scalar  or single fermion times the
  number of degrees of freedom.  
  This pattern   is indeed  directly  seen in \rf{200}--\rf{2.17}. 
 Note that  this relation implies that 
 in $d$ dimensions the   maximal power of $q^{-1}$  in $\wz(q)$  in \rf{2.21}   should   be $d-1$.

  \iffa 
The opposite limit $q\ll 1$ is controlled by thermal energy on $S^{1}_{q}\times \mathbb R^{3}$
 \cite{Kutasov:2000td}
and the leading term is expected to be proportional to the number of dynamical degrees of freedom
$\nu$, or equivalent number of scalar fields
in the flat space path integral partition function $Z\sim [\det(-\partial^{2})]^{-\nu/2}$. 
This means that for bosons  
\be
\la{2.21}
\widehat\zeta(q) = -\frac{\nu}{360}\,\frac{1}{q^{3}}+\mc O(1/q^{2}).
\ee
The 
expansion (\ref{2.21}) is readily checked for the examined fields, including CHS fields
with spin $s$ taking into 
account that  $\nu=s(s+1)$ for them.
For fermions, the prefactor is different, but the $q^{-3}$ term is still proportional to 
the number of degrees of freedom explaining for instance the ratio $1:3$ between $\psi$ and $\psi^{(3)}$
in (\ref{2.16}) and (\ref{2.17}), and same will apply to gravitino.
\fi
\iffa 
> On 27 Jun 2017, at 11:21, Arkady Tseytlin <atseytlin@gmail.com> wrote:
>
> what worries me  is that on S1 x H3   the  coeff of order q
> in F   or E_c   is scheme-dependent; in general, E_c is scheme dependent.
> Thus     d_q  zeta(q) |_{q=1}     can be made  zero or non-zero
> changing E_c.  So could that be   some scheme-dependent issue? That
> does not seem so
> on S^4_q  side where  zeta is unambiguous, but E_c on S3 is ambiguous
> and q-term in F on S1 x H3 is ambiguous... as we are subtracting power
> divergences. may be this   is not real ambiguity as this is not a  due
> to log div.  stillÉ
I think this is an issue in S1 x H3. But on the other hand, if we use zeta on S_q
the regularisation in hyperbolic space must be matched to this. How to do seems non trivial.
But this may be a bit implicit. For instance Klebanov was using a regularised expression for
the initial sum_n log(É) and we discovered his formula could be derived by zeta function.
But it is basically ambiguous up to finite parts in the UV subtraction. Plus, comparison with S1 x H3
has the unsettled UV/IR problem.
\fi

\section{Conformal anomaly coefficients from $\wz(q)$}

Having  found $\wz(q)$   for a CFT on $S^4_q$  one should   be able to extract the information about 
the  corresponding  conformal anomaly coefficients  a  and c in \rf{1.2}. 
The a-coefficient  is the one  appearing in the log divergent part of the partition function on $S^4$. It  
is  thus  simply  proportional to $\wz(1)$ as in    \rf{1.9}, 
\be
\la{3.4}
\text{a} =  -\tfrac{1}{4}\,  \widehat\zeta(1)\ .   
\ee
Starting with \rf{200}--\rf{2.17}   we indeed  match  the  known values of the a-coefficient
for the  $\del^2$ and $\del^4$  conformal scalars,   $\del$ and $\del^3$ fermions
  and spin $s=1,2,3, 4$ CHS fields from \rf{1.3}
\be \la{3.1}
\renewcommand{\arraystretch}{1.5}
\begin{array}{ccccccccc}
\toprule
& \varphi & \varphi^{(4)} & \psi & \psi^{(3)} & \text{CHS}_{1} & \text{CHS}_{2} &\ \  \text{CHS}_{3} &\ \ 
\text{CHS}_{4} \\
\midrule
\text{a} \phantom{\quad}& \frac{1}{360} & -\frac{7}{90} & \frac{11}{720} & -\frac{3}{80} & \frac{31}{180}
& \frac{87}{20} & \frac{171}{5} & \frac{1415}{9} \\
\text{c} \phantom{\quad}& \frac{1}{120} & -\frac{1}{15} & \frac{1}{40} & -\frac{1}{120} &
\frac{1}{10} & \frac{199}{30} &\ \  \frac{914}{15}+\frac{2\,r}{3} &\ \  \frac{890}{3}+\frac{14\,r}{3} \\
\bottomrule
\end{array}
\ee
Here we included also  the known values of the c-coefficient  for  the same 
 $s \leq 2$ fields  and  also the $s=3,4$ 
values from \rf{1.5} depending on the a priori unknown parameter $r$.

While  the value of the function $\wz(q)$ at $q=1$  gives  the a-coefficient,  one observes
from \rf{200},\rf{201},\rf{2.17} 
 that  its  first  derivative   vanishes  at $q=1$   for  all  low spin $s =0, \ha, 1$  conformal fields 
($\wz' (q) = {d \ov d q} \wz(q)$)
\be \la{34}
 \wz'_{\varphi} (1) =0 \ , \qquad 
 \wz'_{\varphi^{(4)}} (1) =0 \ , \qquad 
{\wz}'_{\psi} (1) =0 \ , \qquad 
 \wz'_{\psi^{(3)}} (1) =0 \ , \qquad   \wz'_{1} (1) =0\ . \ee
Surprisingly, this is no longer true for   CHS fields   with $s\geq 2$, i.e. $ \wz'_{s} (1)\not=0$.

The second derivative of $\wz(q)$   at $q=1$  is expected to be related to the 
conformal anomaly c-coefficient.
We  propose the  following general   expression   for $\cc$ 
(and also similar  relation for $C_T$ in other dimensions) 
\be
\la{3.7}
\text{c} =  -\tfrac{1}{4}\,\tfrac{d^{2}}{dq^{2}}\big[q\, \widehat\zeta(q)\big]\Big|_{q=1}
=    -\tfrac{1}{4}\, {\wz}''(1)   -   \tfrac{1}{2}  { \wz}'(1)  \ . 
\ee
In the low-spin cases when  the first derivative vanishes \rf{34},   $\cc$ is then given  just  by the 
second derivative term. 
Using the expressions for $\wz(q)$ in \rf{200}--\rf{2.17}  we indeed  reproduce the known values of $\cc$   for  
fields  with   $s\leq 1$ in  \rf{3.1}.

For $s=2$, i.e. the Weyl graviton,    where $\wz'_2(1)=-8$ is no longer zero,     we   get from \rf{3.7} 
precisely the known value $\cc= {199\ov 30}$ \ci{Fradkin:1981iu}. 
The   same  agreement  is found in the case  of  $d=6$  Weyl graviton as will be discussed in section 4 and Appendix \ref{app:weyl}. 

Note that as  follows   from \rf{3.7}  and the general form of  $\wz(q)$ in \rf{1.9},\rf{2.21}   one has 
\be\la{cpc} 
\cc = E_c + {1\ov 240} \nu  -  {1 \ov 2}  p_2 \  , \ee
where $p_2$   is the coefficient  of the $1\ov q^2$   term in $\wz(q)$. 
Interestingly, $p_2=0$   for all  lower-spin fields (see \rf{200},\rf{201},\rf{2.17})   but is non-zero  for  higher 
spin  CHS fields  starting with Weyl graviton (cf. \rf{202}-\rf{2.15}). 

In the case of CHS fields   with $s=3$ and 4
we find  $\cc_s$  in \rf{3.1}  corresponding to the value   of the parameter $r$ in \rf{1.5}  equal to -1, i.e.
\be\la{115} r=-1: \ \ \ \ \ \ \qquad\qquad 
\cc_{3} = \frac{904}{15}, \qquad  \qquad \cc_{4} = 292  \ . 
\ee
  This    provides  a strong evidence that  the correct  value  of  the c-coefficient of the CHS fields 
  is given by  \rf{1.6}. 

\def \SS  {{\cal S}}\def \FF {{\cal F}}

Let us now   compare  \rf{3.4}  and \rf{3.7} with similar relations  for a  and c expected from  the free energy \rf{2.18}, \rf{2.19} 
on $S^1_q \times \HH^3$. 
Let us   first  recall the expression for the  \renyi entropy   in terms of the free energy on 
$S^1_q \times \HH^3$
\be\la{36} 
\SS(q) = \frac{\FF(q)-  q\, \FF(1)}{q-1}   \ .
\ee
Then the  expected expressions  for the a  and c anomaly coefficients are 
 \cite{Perlmutter:2013gua}
\begin{align}
\la{3.10}
\text{a} &= -\tfrac{1}{4}\,\SS(1) \, = \tfrac{1}{4}\,\mc F(1)-\tfrac{1}{4}\,\mc F'(1)\ , \\
\la{310} \text{c}  &=  \tfrac{1}{2}\,\SS'(1) = \tfrac{1}{4}\,\mc F''(1)\ .
\end{align}
Using the relation \rf{2.19}  between $\wz(q)$  for the conformal  theory on $S^4_q$ 
and $\FF(q)$  on $S^1_q \times \HH^3$ we  conclude that in all low-spin cases 
when $\wz'(1) = -\FF'(1) =0$  \rf{34}  the  expressions \rf{3.10} and \rf{310} are indeed equivalent 
to \rf{3.4}   and \rf{3.7}.\foot{The expression  for a-coefficient    in terms of the entanglement 
 entropy $\SS(1)$  is assumed to  incorporate  the required   edge mode contribution 
 adding  a  constant term to $\FF(q)$  (see   \cite{Huang:2014pfa,Donnelly:2016mlc,Wong:2017pdm})
 which is effectively  included  in the  systematic computation of $\wz(q)$ on $S^4_q$.
 Recent   work  \cite{DowkerNew} extends this 
to the case of a conformally
invariant $p$-form  field in $d=2p+2$.  As in the   case of the 
 Maxwell field  in $d=4$    the correct a-coefficient   is found directly from the 
 spectral computation on $S^d_q$, while a  constant shift is needed in the computation on $S^1_q \times \HH^{d-1}$.
This shift is predicted  \cite{DowkerNew} 
 to be minus  the entanglement entropy of a 
conformal $(p-1)$-form field,  in agreement with \cite{Donnelly:2016mlc}.}
 In particular,  for the $s=1$ case the   conformal  anomaly  coefficients 
are reproduced  correctly   in both $S^4_q$   and $S^1 \times \HH^3$   approaches 
(see  also \cite{Beccaria:2017dmw}). 

The first novel case is the $s=2$ Weyl graviton   when $\wz'(1) = -\FF'(1) \not=0$ 
and the relation \rf{310} is to be  replaced by \rf{3.7}.  The consistency of \rf{3.7} 
for all three  $s=2,3,4$   CHS cases  discussed explicitly  above
 provides  a strong  evidence  for its universal   applicability. 
 A  similar expression  is true also 
 in $d=6$   (where it  leads to the correct $C_T\sim \cc_3$  coefficient for the 6d conformal graviton, see 
 section 4 and Appendix \ref{app:weyl})    and  in $d=2$  (see  Appendix \ref{AA}).
  It   would be important  to derive \rf{3.7} 
 in general   using the approach  analogous  to the one in  \cite{Perlmutter:2013gua},
 taking fully into account the special features of stress tensor for higher spin fields. 

Using  the  expected general  structure  \rf{2.21} of $\wz(q)$  with  
  the   expression  \rf{1.4} 
 for  
 $E_c$   for spin $s$ CHS field 
 as well as the  explicit results for  $\wz_s(q)$  with 
$s=1,2,3$  in \rf{201}--\rf{203} 
  it is possible determine  the  general   form of $\wz_s (q)$  
  for any  value of $s$. 
Starting with an ansatz (with $\nu=s(s+1)$)
\be
\la{3.15}
\widehat\zeta_{s}(q) = -\frac{\nu}{360\, q^3}+\frac{p_2(\nu) }{q^{2}}
+\frac{p_1(\nu ) }{q}+  p_0(\nu) -\frac{\nu\,
(18\,\nu^{2}-14 \nu -11)}{360}\,q,
\ee
where  $p_i(\nu) =\nu( k_{i2} \nu^2  + k_{i1} \nu  + k_{i0})$    are cubic  polynomials  in $\nu$ 
(so  that $\wz_s$ is at most cubic in $\nu$  and vanishes for  $\nu=0$ 
 as required to match the   structure of  conformal anomaly coefficients) 
one is able to fix  the 9   unknown 
coefficients $k_{ij}$   by matching  to  the   $s=1,2,3$  expressions  in \rf{201}--\rf{2.15}.
As a result, 
\be\la{3115}
\widehat\zeta_{s}(q) 
= -\frac{\nu}{360\,q^{3}}-\frac{\nu^{2}\,(\nu-2)}{36\,q^{2}}
+\frac{\nu\,(2\,\nu^{2}-5\,\nu-1)}{36\,q}
-\frac{\nu^{2}(2\nu-1)}{36} -\frac{\nu\,
(18\,\nu^{2}-14 \nu -11)}{360}\,q.
\ee
Then a  highly non-trivial  consistency check  is that for $s=4$  and $5$  this  expression reproduces also 
$\wz_4(q)$ in \rf{2.15}  and  $\wz_5(q)$   in   \rf{a6}. 
 Furthermore, applying \rf{3.4} 
we then  match  the  known a-coefficient in \rf{1.3},  while  applying 
\rf{3.7}  we  get the $r=-1$   expression for the c-coefficient in \rf{1.6}.\foot{The analog of \rf{344} here  contains 
$ n_{\rm z} = s\,{\rm k}_{s}-\sum_{s'=0}^{s-1}{\rm k}_{s'}  = \frac{1}{36} \nu^{2}\,(5\,\nu-1)$, generalizing the expressions in \rf{3.6}.}
Note also that 
\be 
\wz_s'(1) = - {1\ov 60} \nu (\nu-2) ( 3 \nu +2) \ , \la{555}
\ee
is a non-zero integer  for all $s>1$   CHS fields   and thus contributes to $\cc$ in \rf{3.7}. 

We observe also that not only the regularized sums of $\aa_s$ and $\cc_s$   but also 
 the sum of  the  full $\wz_s(q)$   functions over all $s=1,2, ...$ 
   vanishes, i.e.
 \be\la{999}
  \sum_{s=1}^\infty e^{- \epsilon (s+ {1\ov 2})} \, \wz_s(q)\Big|_{\epsilon \to 0, \ \rm finite} =0 \ , 
 \ee 
   so that the full CHS theory is one-loop UV finite  on $S^4_q$ space. 
 
 This  implies  also the  vanishing  of the total  free energy on $S^1_q \times \HH^3$  \rf{2.19}  and  thus 
 of   the  associated \renyi   entropy \rf{36}.
This  vanishing   appears to   be    consistent  with the   "topological" nature  of  the CHS  theory
 \cite{Beccaria:2015vaa}.  Similar conclusions are reached  for the massless  higher spin tower in Appendix \ref{JJ}.

Let us note also   that  as the  (one-loop) 
 logarithmic UV divergences   cancel in the full CHS theory, 
  the finite part of the corresponding  partition function $Z$   is scheme-independent. 
  As was  shown in \ci{Beccaria:2015vaa}, $Z=1$   in flat space (assuming the same  regularization as in \rf{999}, in which the total number of degrees of freedom vanishes)
  and also on $S^4$  (which 
  could be expected  given the cancellation of  conformal a-anomalies). 
   One  may  expect that  since $S^4_q$   has a non-zero Weyl tensor  it  is likely that   $Z(S^4_q)$  is a  non-trivial function of $q$. It would be   interesting to computeit using the heat-kernel method in
    Appendix \ref{app:mass}.

\section{Generalization to six dimensions }

Let us now   demonstrate  how  similar computations of $\wz(q)$  
and related  conformal    anomaly coefficients can be performed  in six dimensions. 
In  6d  for a classically Weyl invariant theory    one gets instead of \rf{1.2} 
\be
\label{4.1}  B_6 ={\te {1\ov (4 \pi)^3} }\int d^6 x \sqrt g\ \bar  b_6(x) \ , \qquad \qquad 
\bar  b_6 =  
  -\text{a}\,E_{6}+ \text{c}_{1}\,I_{1}+\text{c}_{2}\,I_{2}+\text{c}_{3}\,I_{3}\   ,  
\ee
where 
$E_{6} =- \ep_6 \ep_6 RRR$ is proportional to  the 6d Euler density 
   and the  3 independent   Weyl invariants   are 
   $I_1= C_{\alpha\mu\nu\beta}\,C^{\mu\rho\sigma\nu}\,C\indices{_{\rho}^{\alpha\beta}_{\sigma}}$, $I_2 = C\indices{_{\alpha\beta}^{\mu\nu}}\,C\indices{_{\mu\nu}^{\rho\sigma}}\,
 C\indices{_{\rho\sigma}^{\alpha\beta}}$ and $I_3
 = C_{\mu\alpha\beta\gamma}\, D^{2} C^{\mu\alpha\beta\gamma} + ...$
(see for details \cite{Bastianelli:2000hi} and Appendix~\ref{app:weyl}). 

The aim   will be to  consider  the  conical deformation   $S^6_q$ of  6-sphere  (with the metric as in \rf{1.7} with $S^2$ singularity replaced by $S^4$  one),  
compute the spectral   $\zeta$-function at $z=0$ or $\widehat\zeta(q)=B_6$  as in \rf{1.8}
and then  extract the values of  the conformal anomaly coefficients  from it. 
As  the  log divergent part of the free energy on $S^6$   should be  proportional to $\aa$, 
we should have again $\wz(1)\sim  \aa$. 
The  $\cc_3$  coefficient  proportional to $C_T$ in   the 2-point function  of stress tensors  should   be 
determined, as in 4d  case, 
 by the 2nd derivative  of $\wz(q)$ at $q=1$. 
  The coefficients $\cc_3$ and $\cc_4$  related to 3-point 
 functions of stress tensor  may be  possible  to extract from the 3rd (or higher)  derivative of 
 $\wz(q)$   but we will not attempt this here. 
 
Taking into account  normalizations, the expected relations are then the  direct analogs of \rf{3.4} and \rf{3.7} in 4d case:
\begin{align}
\text{a}&\te = -\frac{1}{96}\,\widehat\zeta(1) \ ,  \la{62} \\
\text{c}_3 &\te =  \frac{1}{12}{ d^2\ov d q^{2}} \big[q\,\widehat\zeta(q)]\Big|_{q=1} =   
 \frac{1}{12}  \wz'' (1) + \frac{1}{6}  \wz'(1)  \ . \la{63}
\end{align}
The  bosonic  totally symmetric  rank $s$ conformal  higher spins in 6d  have  kinetic terms $
h_{s}\,\Box^{s+{d-4 \over 2} }\, h_{s} = h_{s}\,\Box^{s+1 }\, h_{s}$. Below 
we shall consider  only  the lowest spin cases:
$s=0$ --  the standard  $\del^2$ conformal scalar, $s=1$ --  the higher derivative 
$\del^4$ vector  \ci{Beccaria:2015uta,Beccaria:2015ypa,Beccaria:2017dmw} 
  and  $s=2$  --  the $\del^6$   conformal graviton  (see Appendix \ref{app:weyl}). 
The corresponding    partition functions   on $S^6$ are \ci{Tseytlin:2013fca}  (cf. \rf{22},\rf{23})
 \begin{align} \la{422} 
&Z_{0}=  \Big[\frac{1}{\det\Delta_{0}(6)}\Big]^{1/2} 
 \ , \qquad\qquad  \qquad  Z_{1}= \Big[\frac{\det\Delta_{0}(0)}{\det\Delta_{1\perp}(7)\ \det\Delta_{1\perp}(5)} 
\Big]^{1/2}  \ , \\
&Z_2 =  \Big[\frac{\det\Delta_{1\perp}(-5)\,\det\Delta_{0}(-6)}
{\det\Delta_{2\perp}(8)\,\det\Delta_{2\perp}(6) \,\det\Delta_{2\perp}(2)}
\Big]^{1/2} \ ,  \la{423} 
\end{align}
where $\Delta_{s\,\perp} (M^2) =( - D^2 + M^2)_{s\,\perp} $  are defined  on $S^6$. 
Assuming as in 4d case that the  these partition functions   have the same structure  on $S^6_q$, 
their computation  requires the knowledge of the spectrum of $\Delta_{s\,\perp} (M^2)$ on this 
 space.  
 
 The analysis of the spectrum  goes along the same lines  as  in Section 2. 
 The  analogs of the  eigenvalues in 
(\ref{2.1}) and (\ref{2.2}) are   obtained  after the replacement 
  $n+\gamma m+3\to n+\gamma m+5$. 
  The  degeneracies  of the 
spectrum for  the conformal scalar and the  4-derivative spin 1 field   are found  from 
  (\ref{2.7}) and (\ref{2.8}) where now $d\to 6$.
The spin 2 degeneracies turn out to be  ($n+m \ge 2 $)
\begin{align}\la{424}
& \text{d}^{(2)}_{n,0} = {\tfrac{1}{12}\, (n-1)(n+2)(n+3)(7 n+22)},\quad
& \text{d}_{n,1}^{(2)} = {\tfrac{1}{6}\, n (3+n)(4+n)(7n+17)},\notag \\
& \text{d}^{(2)}_{n,m>1} = \tfrac{7}{6}\,(n+1)(n+2)(n+3)(n+4).
\end{align}
As a result,  $\widehat\zeta_{s\,\perp}(q)$ in \rf{1.8}   for the operators $\Delta_{s\, \perp}(M^2)$  on $S^6_q$  are given by (cf. \rf{288})
\begin{align}
\widehat\zeta_{0}(q; M^{2}) & \te = \frac{1}{15120\,q^{5}}+\frac{6\,M^{2}-35}{4320\,q^{3}}
+\frac{24-10\,M^{2}+M^{4}}{144\,q} 
 +\frac{4315-3990\,M^{2}+1050\,M^{4}-84\,M^{6}}{30240}\,
\,q, \notag \\
\widehat\zeta_{1\,\perp}(q; M^{2}) &=\te  
\frac{1}{3024\,q^{5}}+\frac{6\,M^{2}-41}{864\,q^{3}}+\frac{5\,(35-12\,M^{2}+M^{4})}{144\,q}\notag\\
& \quad \te 
-\frac{553-210\,M^{2}+15\,M^{4}}{180}\,  +\frac{9439-6342\,M^{2}+1302\,M^{4}-84\,M^{6}}{6048}\,\,q,\la{444} \\
\widehat\zeta_{2\,\perp}(q; M^{2}) &= \te \frac{1}{1080\,q^{5}}-\frac{1}{6\,q^{4}}
+\frac{1111+42\,M^{2}}{2160\,q^{3}}+\frac{6M^{2}-{55}}{6q^{2}}
 +\frac{1560-242\,M^{2}+7\,M^{4}}{72\,q}\notag \\
&\te\quad  -   \frac{3166 - 1860 M^2 +    105 M^4      }{180}    +\frac{  17167-9198\,M^{2}+1554\,M^{4}-84\,M^{6}       }{2160}\,q. \notag
\end{align}
Forming the combinations  of these  functions corresponding to the partition functions \rf{422},\rf{423} 
taking into account as in \rf{344}   the  zero mode contributions\foot{The number of zero modes associated with a  transverse traceless totally symmetric rank  $s$ field on $S^6$  is (see  \cite{Tseytlin:2013fca}  and refs. there)
$\ \ \ 
\text{k}_{s} = \tfrac{1}{4320}(2s+3)s(s+1)^{3}(s+2)^{3}(s+3) $, i.e.  $
\text{k}_{0} =0, \ \text{k}_{1} =1, \ \text{k}_{2} =28, $  etc.}
\be
s=0: \ \ n_{\rm z} = 0; \qquad
s=1: \ \ n_{\rm z} = 2\,{\rm k}_{1} = 2; \qquad
s=2: \ \ n_{\rm z} = 3\,{\rm k}_{2}-{\rm k}_{1} = 83 \ , 
\ee
we  find that the total coefficients $\widehat\zeta_s(q)$  
of the  log UV divergence  of the  CHS partition  functions  \rf{422},\rf{423}  on $S^6_q$ are given by 
(cf. \rf{200}--\rf{202})
\begin{align}
\la{4.3}
 \widehat\zeta_{0}(q) = &
\frac{1}{15120\, q^5} +\frac{1}{4320\, q^3}\,+\frac{31}{30240}\,q,\\
\la{43}  \widehat\zeta_{1}(q)  = &
\frac{1}{1680\, q^5}\,-\frac{1}{288\, q^3 }\,
-\frac{1}{6\, q}\, -\frac{14}{15} 
       -\frac{39}{224}\,q\\
\la{433}
\wz_2(q)= & 
\frac{1}{420\, q^5}\,-\frac{1}{2\, q^4}\,+\frac{703}{360\, q^3}\,
-\frac{23}{2\, q^2}\,+\frac{49}{3\, q}\,-\frac{181}{9}
-\frac{4143}{280}\,q\ .
\end{align} 
As in the  4d case \rf{2.21}  the $q\to 0$ and $q\to \infty$   asymptotics of $\wz(q)$ 
are controlled by free  energies on $S^1_q \times \mathbb R^5$   and $\mathbb R\times S^5$ 
respectively,  i.e. 
\be\la{455}
\widehat\zeta_{s}(q) = \frac{\nu}{15120}\,\frac{1}{q^{5}}+...  + (-2\,E_{c})\,q\ ,
\ee
where the  number of dynamical  degrees of freedom 
 $\nu$   \ci{Tseytlin:2013fca} and the Casimir energy on
$\mathbb R\times S^{5}$ \cite{Beccaria:2014jxa} for 6d CHS fields 
are  given by 
\begin{align}
\la{4.5}
&\nu=\tfrac{1}{4}(s+1)^{2}(s+2)^{2}, \ \ \ \ \ \ \ \qquad \ \ \    s=0,1,2, ...
\\
& \E_{c, s} = \tfrac{1}{60480}\,\nu\,(
96\,\nu^{3/2}-232\nu-12\nu^{1/2}+117)\ .\la{454}
\end{align}
The values of $\wz_s$ at $q=1$     reproduce   \rf{62}  the  known 
  $\aa$-coefficients  
  \cite{Tseytlin:2013fca}\foot{With  our normalization in  the 6d conformal scalar   case ($\nu=1$)   we get 
  $\aa_0= -{1\ov 72576}$.}
\begin{align}
\la{4.6}
\text{a}_{s} = \tfrac{1}{1814400}\,\nu\,(88\,\nu^{3/2}-110\nu-4\nu^{1/2}+1)  \ . 
\end{align}
We  also observe that as in the 4d case  \rf{34},\rf{555}    the first derivative $\wz'(1)$  vanishes   for $s=0$ and $s=1$  but not for
$s=2$:
 \be\la{466}
   \widehat\zeta_0'(1)=0 \ ,\ \ \  \qquad \wz_1'(1) =0\ , \qquad \ \ \ 
 \wz_2'(1) = -12 \ . 
 \ee
Using \rf{63}      we get  the following values for the $\cc_{3, s}$  coefficients 
\be \la{477}
\cc_{3,0}=   \frac{1}{2520}  \, \qquad \qquad \cc_{3,1}=  -\frac{5}{168}  \, \qquad \qquad 
\cc_{3,2}=    -\frac{1639}{420} \ .
\ee 
The $s=0$   and $s=1$  values  match the  known ones found  earlier  in 
  \cite{Bastianelli:2000hi,Beccaria:2017dmw}.
Remarkably, the   $s=2$   result for $\cc_{3,2}$    agrees  
  with the direct   computation  of the corresponding 
  6d Seeley-DeWitt coefficient  for the 6d Weyl graviton 
  that we present  in Appendix \ref{app:weyl}  where we also  determine   the 
   values of the two other  conformal anomaly coefficients $\cc_1$ and $\cc_2$ in \rf{4.1}
   (see \rf{A11}).
   This  provides a   non-trivial  check of the consistency of the relation \rf{63} for  the
    $C_T \sim \cc_3$   coefficient in 6d. 
   
   For completeness,  let us  summarize the values  of the conformal anomalies for the 6d $s=0,1,2$ CHS fields   below:
\be
\la{4.9}
\renewcommand{\arraystretch}{1.5}
\begin{array}{cccccc}
\toprule 
{s} & \text{a} &  \text{c}_{1} & \text{c}_{2} & \text{c}_{3} 
 \\
\midrule
0 & -\frac{5}{72\times 7!} & -\frac{1}{540} & \frac{1}{3024} & \frac{1}{2520} 
\\
1 & \frac{275}{8\times 7!} & \frac{97}{180} & \frac{911}{5040} & -\frac{5}{168} 
\\
2 & \frac{3005}{2\times 7!} & \frac{1507}{45} & \frac{635}{126} & -\frac{1639}{420} 
 \\
\bottomrule
\end{array}
\ee


\iffa 
\section{Comments} 
\begin{itemize}
\item The ghost terms involve the terms
$(n+\gamma m-2)(n+\gamma m+5)$. This means that there are negative modes around $\gamma=1$
when $n+m<3$. This is possible for all three determinants with spin $<3$. However,
this is a problem concerning 
a finite number of terms, so it should alter $\zeta$ by a $\gamma$ independent constant and thus be 
irrelevant.

\item What about Killing tensors, i.e. low spin modes that do not contribute when
used to express the spin 3 field in terms of $3\perp$ ? But again, this involves a finite number of modes. 

\item About difficulties with smoothed cones: 
one subtle issue that may be related to UV vs IR and cutoffs is
that on cone  if we regularize it we have some epsilon parameter, but
we also have UV cutoff $\Lambda$,  and  when we  we look  at eff action it is
$G(\eps, \Lambda)$   and  order of limits may matter:   if we first   expand at
large $\Lambda$ we get:    power div  $\Lambda^n$  + zeta(eps)  log $\Lambda$ +  finite
1/$\Lambda^k$  terms
then we take   limit $\eps\to 0$; but taking $\eps\to 0$
first and then $\Lambda\to\infty$
not give same result -- both $\eps$   and $1/\Lambda$ to zero are short-distance
limits and their  order may matter...   This is not a constructive
comment, I am just worrying  that looking at eps-regularized Seeley
coeffs may not make much sense   and a reflection of that may be
$1/\eps^2$ singular terms appearing...

\item  Is method of images  in sec. 5 of \cite{Dowker:1988pp} useful ? It seems not because his cone
is flat. Too simple case. 

\item by the way,  if one could  assume CHS factorization on Einstein space
then we could  fix   both a  and c   in one shot -- as we realized
for spin 1 vector
in recent paper.  Now, we  know that on a sphere CHS does not
factorize   into same   operators but rather into a combination with
different masses. At the same time,   on Ricci flat space   we had for
s=2  the same   2 operators and I boldly assumed  that this pattern
continues to s>2.   But what if that starting with s=3  the  s
Lichnerowitz-type operators  are not same
--  may   differ  by coeff  of $C_{mnkl}$ term ?    Unless that
contradicts something that would explain  a possibility to get r=-1
instead of r=1/2  result ...  But of course naively factorization is
already ruled out by  Nutma-Taronna...
Reason for this suggestion is that on the cone  $S^4_q$ we  start with
ops on sphere that do have different massess  (partially massless etc)
and may be via cone that  effectively translates also to different
coeffs of $C_{mnkl}$  terms on general Einstein space ...

\item About Perlmutter derivation. Similar logic should be posisble directly on 
$S_{q}^{4}$. Main  issue  is what distinguishes $s<2$  from $s=2,3, \dots$  CHS.
According to App. D of \cite{Beccaria:2017nco}
 for spin $s>1$ it is necessary 
to use the equations of motion to prove gauge invariance of the improved
symmetric traceless stress tensor. This could be related to a non zero value
of $\langle T\rangle$ that is the same as $\widehat\zeta'(1)\neq 0$. This can be 
an indication why $s>1$ cases are different from the analysis in Perlmutter.

\end{itemize}

\fi 

\iffa 
\medskip
\begin{verbatim}
end of intro:   after 1.8: here we may say that    instead of trying
to apply  general relation 1.2 for conf anomaly here we may use defn
of c as   coeff  in <TT> in near flat space expansion -- by analog y
with Perlmutter.
Systematic way may be as follows: if we start with usual sphere metric
in cartesian coordinates
ds^2 =  dx_m dx_m /(1 + x^2/(4r^2))^2
where  r is curvature radius   the near-flat-space expansion is
obvious -- ds^2 =   dx_m dx_m -  1/2  x^2/r^2 dxdx + ...
If we could rewrite  1.3 for  q   near  1   in a similar way that
will give us similar     starting point:
rescaling tau  by q  we get:

ds^2(q)  = ds^2(S4) + (q^2-1)  cos^2 theta dtau^2  =
where tau=(0, 2pi)
then we may  express theta and tau in terms of above x_m coordinates.
Then we may  expand  eff action in  insertions of
(q^2-1)  h_mn  dx^m dx^n = (q^2-1)  cos^2 theta dtau^2
which will be  expansion  in correlators of T_mn
in S4  metric:
Gamma[S4_q] = Gamma[S4]

+ (q^2-1)  <T_mn>  x h_mn +  (q^2-1)^2  <T_mn T_pq > x  h_mn  x  h_pq
+ ...                (*)

where    x  stands for integration over  S4

But Gamma[S4_q] = - zeta(q) log L +  finite

so  we need to extract log L term from (*) -- i.e. singular parts of
integrals there.  One this is done it is clear that

zeta'(1)   ~   2  <T_mn>  x h_mn
zeta"(1) ~   2  <T_mn>  x h_mn  +  8  <T_mn T_pq > x  h_mn  x  h_pq

as  <T_mn T_pq >  ~    c_T
that suggests that   c_T should be related to a combination of zeta"
and zeta'  in general   (if zeta'  is non-zero).


what I find is very surprising is that ration of zeta"  and zeta" in
c_T is same in 4d and 6d, i.e.
c_T = k_d ( q z')' (q=1)
where k_d depends on dimension.
That suggests some  universality in the derivation.

may be  above or  below 1.8  it should be said that it  is not sepcific to 4d
\end{verbatim}
\fi

\section*{Acknowledgments}
We would like  to thank 
S. Dowker  and   D.  Fursaev    for useful  discussions. 
The work of AAT  was   supported by the ERC Advanced grant no. 290456,
 the  STFC Consolidated grant ST/L00044X/1  and   the Russian Science Foundation grant 14-42-00047 at Lebedev Institute.


\appendix


\section{Details of   computation of $\widehat\zeta(q)$   for  4d massive scalar
\la{app:mass}}

Here   we provide some details of the computation of $\wz(q)$   in section 2 on the example of 4d scalar operator $\Delta_{0}(M^{2})=-D^2 + M^2$.  The case of the  conformal  coupling  on unit-radius $S^4$  corresponds to $M^2=2$. 
Introducing the parameter $\mu$   related to $M^2$ by  
$M^{2} = \frac{9-\mu^{2}}{4}$
and   setting  $\gamma=1/q$   we then  have for the  corresponding   spectral $\zeta$-function in  \rf{1.8}
\be
\la{P.1}
\zeta(z; \gamma) =\sum_{n,m=0}^{\infty}\text{d}_{n,m}^{(0)}\, 
\big [(\tfrac{3}{2}+n+m\,\gamma)^{2}-\tfrac{\mu^{2}}{4}\big]^{-z} \ . 
\ee
We  used \rf{2.1} and   \rf{2.7}, i.e.  $\text{d}^{(0)}_{n,0} = \frac{1}{2}\,(n+1)\,(n+2)$, \  
$\text{d}_{n,m>0}^{(0)} = (n+1)\,(n+2)$. 
The evaluation of (\ref{P.1}) was  first considered in  \cite{Fursaev:1993hm} where it  allowed to 
obtain the finite-temperature one-loop effective potential for a scalar field in 
 de Sitter space-time. 
The result 
revealed an unexpected  dependence
of the  logarithmic divergences  on the temperature associated  to the presence of a horizon
which is directly related to $q$-dependence of $\wz(q)=\zeta(0;q)$   we discussed in section 2.\footnote{In the present context  the parameter $\gamma=q^{-1}$  of the  conical 
singularity    corresponds to the  ratio of  the temperature  and  the Hawking temperature  in   \cite{Fursaev:1993hm}.}

We will   compute   $\zeta(0; \gamma)$ in (\ref{P.1}) by using a
somewhat  more direct  method than employed in \cite{Fursaev:1993hm}.  
We first split the contribution from the 
$m=0$ and $m>0$ modes as 
\begin{align}
&\zeta(z; q) = \zeta^{(a)}(z; q) + \zeta^{(b)}(z; q), \quad \qquad 
\notag 
\zeta^{(a)}(z; q) = \sum_{n=0}^{\infty}\tfrac{1}{2}(n+1)\,(n+2)\,[(\tfrac{3}{2}+n)^{2}-\tfrac{\mu^{2}}{4}]^{-z}, \\
\la{P2}
&\qquad\qquad 
 \zeta^{(b)}(z; \gamma) = \sum_{n=0}^{\infty}\sum_{m=1}^{\infty}(n+1)(n+2)\,\big[
 (n+m\,\gamma+\tfrac{3-\mu}{2})(n+m\,\gamma+\tfrac{3+\mu}{2})\big]^{-z}.
\end{align}
As    we are interested only  in the value at $z=0$    each of the two  terms  $\zeta^{(a)}$  and $\zeta^{(b)}$   can be computed by expanding 
to  quadratic order in  $\mu^{2}$ only  (higher order terms  in $\mu$   will   give
 vanishing contributions  in the $z\to 0$ limit, cf.  also  footnote \ref{foot:mu}).   
 For the  first term we get   
\begin{align}
\la{P.4}
\zeta^{(a)}(z; q) =&
\tfrac{1}{8} \left[4 \zr \left(2 z-2,\tfrac{3}{2}\right)-\zr \left(2
   z,\tfrac{3}{2}\right)\right]+\tfrac{1}{32} \mu ^2 z \left[4 \zr \left(2 z,\tfrac{3}{2}\right)-\zr
   \left(2 z+2,\tfrac{3}{2}\right)\right]\notag \\
   & +\tfrac{1}{256} \mu ^4 z (z+1) \left[4 \zr \left(2
   z+2,\tfrac{3}{2}\right)-\zr \left(2 z+4,\tfrac{3}{2}\right)\right]+\mc O(\mu ^6) \ , 
   \end{align}
   where $\zr(a,b)$ is the Hurwitz zeta-function \rf{72}.
Dropping the  contributions that manifestly vanish at  $z\to 0$ 
(due to explicit factors of $z$ that multiply analytic terms) we  find
\be
\la{P.5}
\zeta^{(a)}(z; q) =
\tfrac{1}{8} \left[4\, \zr \left(2 z-2,\tfrac{3}{2}\right)-\zr \left(2
   z,\tfrac{3}{2}\right)\right]+\mc O(z) \ . 
\ee
This  vanishes  at $z=0$
\be
\la{P.6}
\zeta^{(a)}(0; q) = \tfrac{1}{8}\,
  \big[4\times(-\tfrac{1}{4})-(-1)\big] = 0.
\ee
Instead of following   the same strategy in the case of $\zeta^{(b)}(0; q)$
we shall  use    a simpler  approach   by relating it to the  $t^0$ coefficient in the 
expansion of the corresponding heat kernel replaced   by the sum 
of the kernels  corresponding to the "1-st order" factors in \rf{P2}\foot{As the eigenvalues factorize, the same applies to the corresponding determinant, and thus the  heat kernel can be replaced by a sum of 
heat kernels corresponding to the factors. 
The origin of the   $1\ov2$
 factor 
 in the  relation for $\zeta(0;q)$  may be understood  by  comparing dimensions of the proper-time   cutoffs
in the original heat kernel and its "factor" analogs, cf.  also 
 \cite{Allen:1983dg}.
}
\begin{align}
K(t;\gamma) = & \,\sum_{n=0}^{\infty}\sum_{m=1}^{\infty}(n+1)(n+2)\,\Big[
e^{-t\,(n+m\,\gamma+\frac{3-\mu}{2})}+e^{-t\,(n+m\,\gamma+\frac{3+\mu}{2})}\Big]\notag \\
&  \stackrel{
t\to 0}{\sim}
\sum_{k=-4}^{\infty}\,  h_{k}\,t^{k} \ , \ \ \ \ \ \ \ \qquad  \    \zeta^{(b)}(0; q) = \tfrac{1}{2} h_{0} \ . \la{AA6}
\end{align}
Computing  the two sums, we readily obtain
\begin{align}
K(t; \gamma) = &\frac{e^{-\frac{1}{2} (\mu -3) t} \left(e^{\mu  t}+1\right)}{\left(e^t-1\right)^3 \left(e^{\gamma
   t}-1\right)} = \frac{2}{\gamma }\frac{1}{t^{4}}-\frac{1}{t^{3}}+\Big(
   \frac{\mu ^2-1}{4 \gamma }+\frac{\gamma }{6}\Big)\,\frac{1}{t^{2}}+\frac{1}{8} \left(1-\mu
   ^2\right)\,\frac{1}{t}\notag \\
   & -\frac{\gamma ^3}{360}+\frac{\mu^2 -1 }{48} \gamma  +\frac{5 \mu ^4-30 \mu
   ^2+17}{960 \gamma }+\mc O(t)\ . \la{AA8}
\end{align}
The $t^0$  term here gives 
 (using \rf{P.6}  and $\gamma=1/q$)
 \be
 \zeta(0; q)= \zeta^{(a)}(0; q) + \zeta^{(b)}(0; q) =-\frac{1}{360\, q^3}
 +\frac{\mu^2-1}{48\, q}\,
 +
  \frac{5\,\mu^{4} -30\,\mu^{2}+17}{960}\,q
\ , \la{AA9}  \ee
which is equivalent to the expression in \rf{288}  after we recall that  $\mu^2= 9 - 4 M^2 $.
For example, for the  conformally coupled scalar   with $\mu=1$  we get, in agreement with \rf{200}, 
 \be\la{AA10}
 \zeta(0; q) = -\frac{1}{360\,q^{3}}-\frac{q}{120}.
 \ee

\iffa 
Here  $\mc O(\mu^6) $  terms do not contribute in the limit  $z\to 0$.
The same should  also  apply to extra   $\mc O(\gamma^{4})$   terms:
 the highest  power of
 $\gamma$   in  free energy on $S^d_{1/\gamma}$ should be  $\gamma^{d-1}$ 
 as  in the   $\gamma \to \infty$ limit   we should  match  the high-temperature limit of the standard  thermodynamic partion function 
 on $S^1_{1/\gamma} \times  \mathbb R^{d-1}$ (see  section 3).  
\fi

\section{Degeneracies  of eigenvalues of  bosonic spin $s$ Laplacian on  $S^4_q$}
\la{app:den}

The degeneracies for $\gamma\in[1,2)$ for spin $s\le 4$  in \rf{2.8}--\rf{282}
admit a natural generalization to all integer $s>0$  (here  $n+m \ge s $)
\begin{align}
\la{M.1}
& \text{d}^{(s)}_{n,0} = \tfrac{1}{2}\, (n-s+1)\,[(2s+1)\, n + K_0], \qquad
 \text{d}^{(s)}_{n,1} = (n-s+2)\,[(2s+1)\, n + K_1], \notag\\
 &\text{d}^{(s)}_{n,2} = (n-s+3)\,[(2s+1)\, n + K_2], \qquad ...\notag\\
& \text{d}^{(s)}_{n,s-1} = n\,[(2s+1)\, n + K_{s-1}], \qquad 
 \text{d}^{(s)}_{n,m>s-1} = (2s+1)\,(n+1)(n+2) \ , 
\end{align}
  where the integers $K_{p}$  are 
\be\la{M1}
K_p = 2\,(1+2\,p)+(3+2\,p)\,(s-p),\qquad \qquad  p=0,\dots, s-1.
\ee
One can check  that the degeneracies in (\ref{M.1}) are always non-negative and also even
when $m>0$ as follows from the expected symmetry of the  spin $s$ Laplacian eigenstates under
the exchange $x^{+}\leftrightarrow x^{-}$ in this case. 
The total degeneracy
\be
\sum_{m=0}^{s-1}\text{d}^{(s)}_{N-m,m}+\sum_{m=s}^{N}\text{d}^{(s)}_{N-m,m} =
\tfrac{1}{6}\,(2s+1)\,(2N+3)(N+s+2)(N-s+1)
\ee
is equal  as it should  to the  degeneracy of the level $N$ eigenvalue  for the regular 
 sphere $S^{4}$  (cf. footnote \ref{foot:tot}).
 
 A further test of (\ref{M.1})
is provided by the explicit calculation of the  zeta function $\wz_s (q)$  for the CHS field with
spin s. For instance,  
 for $s=5$
(\ref{M.1})   gives  ($n+m \ge 5 $)
\begin{align}
\la{M.4}
& \text{d}^{(5)}_{n,0} = \tfrac{1}{2}\,(n-4)\,(11\,n+17), \qquad
 \text{d}^{(5)}_{n,1} = (n-3)\,(11\,n+26), \notag \\
& \text{d}^{(5)}_{n,2} = (n-2)\,(11\,n+31), \qquad\ \ \ 
 \text{d}^{(5)}_{n,3} = (n-1)\,(11\,n+32), \notag \\
&  \text{d}^{(5)}_{n, 4} = n\,(11\,n+29), \qquad\qquad\quad \
  \text{d}^{(5)}_{n,m>4} = 11\,(n+1)\,(n+2).
\end{align}
Generalizing the   spin 5  CHS partition function  on $S^4$  \ci{Tseytlin:2013jya}
\be\la{a5}
Z_{5}
= \Big[
\frac{\det\Delta_{4\perp}(-24)\det\Delta_{3\perp}(-25)\det\Delta_{2\perp}(-26)
\det\Delta_{1\perp}(-27)\det\Delta_{0}(-28)}{
\det\Delta_{5\perp}(7)\det\Delta_{5\perp}(5)\det\Delta_{5\perp}(1)
\det\Delta_{5\perp}(-5)\det\Delta_{5\perp}(-13)
}
\Big]^{1/2},
\ee
to $S^4_q$   we find that 
 the associated  $\widehat\zeta_{5}(q)$  function determined using  (\ref{M.4}) is given by 
\begin{align}
\widehat\zeta_{5}(q) &= -\frac{1}{12 q^3}-\frac{700}{q^{2}}+\frac{8245}{6\,q}-1475
-\frac{15769}{12}\,q \ . \la{a6}
\end{align}
This  expression is in perfect agreement 
with our general proposal in  \rf{3115}  (here $\nu= 30$).

\section{Two-dimensional  case \la{AA} }

In two dimensions there  is  just the ``a'' coefficient of the Weyl    anomaly 
  that also  has the  interpretation of  the coefficient $C_T$ in the 2-point function  of stress tensor, 
i.e.  is the Virasoro central charge
and thus is usually denoted as $c$. In standard normalization where a 
real $\del^2$  scalar has $c=1$   we have  for the  coefficient of the  log UV divergence of the partition function (cf. \rf{1.1})
\be  \la{ap1} 
B_2 = {1\ov 4 \pi}  \int  d^2x \sqrt g \ \overline b_2 \ , \ \ \ \ \qquad 
\overline b_2 = \aa \, R \  , \qquad \aa\equiv  \tfrac{ 1}{ 6} \, c \ . \ee
On $S^2$   one  thus finds  $B_2= \tfrac{1}{3}   c$. 
For  a  conformal field defined  on a conical deformation  $S^2_q$   of  the  2-sphere 
we expect the  corresponding spectral zeta-function  at $z=0$ 
 to  have a   similar general form as in 4d \rf{2.21}  and in 6d \rf{455}, i.e. 
\be
\la{N.1}
\widehat\zeta(q) = \frac{\nu}{6\, q}+p_{0} -2\,E_{c}\,q,
\ee
where the  first and the  last terms are fixed   by the asymptotics corresponding to 
$S^1_q  \times \mathbb  R $ ($q\to 0$)    and $ \mathbb  R \times S^1$   ($q\to \infty$). 
Here $\nu$  is the  number of effective degrees of freedom with $\nu=1$ for a real $\del^2$ scalar
and $\nu= -2$   for  the 2d conformal higher spin fields   with kinetic terms 
$h_s  \Box^{s + {d-4}\ov 2}  h_s = h_s \Box^{s-1} h_s$   with $s=1,2, ...$  \cite{Tseytlin:2013fca}.
The  Casimir energy  $E_c$   on  unit-radius 
$ S^1$  
 should  in general  be related to  the  central charge $c$
by \ci{Bloete:1986qm,Herzog:2013ed}  $E_c= - \tfrac{1}{12}  c$. 
Since  $B_2(S^2) =  \wz(1)$  we then conclude that 
\be   c = 3\, \wz(1) = -12 E_c \la{ap2} \ . \ee
As follows from \rf{N.1}  and $E_c= - \tfrac{1}{12}  c$   we then have  also 
the following representation for $c$ 
\be 
c = 3\, \tfrac{ d^2}{ d q^2} \big[ q\, \wz(q)\big]\Big|_{q=1}   \ . \la{ap5} \ee
Remarkably, this  relation for $c=C_T$ in 2d case  is a direct counterpart 
of the similar  4d  \rf{3.7} and 6d \rf{63} relations we proposed above. This  supports their  
common origin   and  implies a  universal  applicability  of $C_T \sim   \wz"(1) + 2 \wz'(1)$  relation 
in any dimension.

In the case  of  the standard  $\del^2$   scalar field   \rf{N.1},\rf{ap2} imply that 
 \be
\la{N.3}
\widehat\zeta(q) ={1\ov6\,  q}  +  \frac{1}{6}\, q  \   .
\ee
In the  case of $d=2$ CHS fields one finds \cite{Tseytlin:2013fca,Giombi:2013yva,Giombi:2014iua}
\be\la{ap3} 
c_{s} = -2\,\big[1+6\,s(s-1)\big]\ , \ \ \ \qquad \ \ \  s\geq 2 \ . 
\ee
As a result,  the function 
 $\wz(q)$   for the 2d  CHS fields  consistent   with the above relations \rf{ap2},\rf{ap3} 
 turns out to be 
\be\la{ap8}
\widehat\zeta_{s}(q) = -\frac{1}{3\,q}-2\,s\,(s-1)-\frac{1}{3}\,\big[1+6\,s(s-1)\big]\,q=
-\frac{1}{3}\big({1\ov q}-1\big)  + {c_s \ov 6} \,\big( q +1\big) 
\ .
\ee
Note that  while  for  a standard scalar \rf{N.3} we find  that $\wz'(1)=0$, for the CHS fields 
$\widehat\zeta'_{s}(1) = -2\,s\,(s-1)$  so  that   for $s\geq 2$  it is  again  non-zero as in 
the 4d and 6d  cases.


\section{
Conformal anomaly coefficients for the  Weyl graviton in six dimensions}
\la{app:weyl}

 \def\bb {\bar b}\def \D {\Delta} \def \d {\delta} \def \F  {{\cal F}}

In 6d there are  three  dimension 6  non-trivial  
Weyl invariants  $I_1,I_2,I_3$  that appear in (\ref{4.1}) 
 \begin{align}
 I_{1} &= C_{\alpha\mu\nu\beta}\,C^{\mu\rho\sigma\nu}\,C\indices{_{\rho}^{\alpha\beta}_{\sigma}},
 \qquad I_{2} = C\indices{_{\alpha\beta}^{\mu\nu}}\,C\indices{_{\mu\nu}^{\rho\sigma}}\,
 C\indices{_{\rho\sigma}^{\alpha\beta}}, \notag \\
 I_{3} &= C_{\mu\alpha\beta\gamma}\,(D^{2}\,\delta^{\mu}_{\nu}+4\,R^{\mu}_{\nu}-
 \tfrac{6}{5}\,R\,\delta^{\mu}_{\nu})\,C^{\nu\alpha\beta\gamma}+\text{total derivatives}.\la{B.1}
 \end{align}
 A candidate  Weyl-invariant gravity   action  is then an integral of  a   linear combination 
 of these  3 invariants. 
 
  There is a particular   choice  \ci{Bonora:1985cq}
 \ \ $\mc W_{6} =- I_3  +  3 I_{2} +12 I_1$   that  has    special 
 properties: (i) it vanishes  on a Ricci flat background, and  (ii)  it admits $(2,0)$  
locally   superconformal  extension \ci{Beccaria:2015uta,Butter:2017jqu,Butter:2016qkx}.
Related to (i)   and (ii)      is that  $\mc W_{6} $  
 appears, respectively,  as  the coefficient of the  logarithmic IR 
   divergence of the Einstein action in AdS$_7$   evaluated on the solution of Dirichlet problem \ci{Henningson:1998gx} and also 
   as the  log UV divergence  of  the $(2,0)$  tensor multiplet   \cite{Bastianelli:2000hi}.
 The resulting action may be written as 
  \be
 \la{B.2}
S = \int d^{6}x\,\sqrt{g}\,\Big[R^{\mu\nu}D^{2}R_{\mu\nu}
-\tfrac{3}{10}\,R\,D^{2}\,R-2\,R^{\mu\nu\rho\sigma}\,R_{\nu\rho}\,R_{\mu\sigma}
-R^{\mu\nu}R_{\mu\nu}\,R+\tfrac{3}{25}\,R^{3}\Big] \ .
\ee
The fact that it is expressed in terms of the Ricci tensor  and is at most linear in the Weyl tensor
  implies  that it 
  can be rewritten  as a 2nd derivative action  involving  several tensors  of rank $\leq 2$
  and it  is uniquely selected by this requirement \ci{Metsaev:2010kp}.

 The quadratic part of  \rf{B.2} expanded around
a curved  space  is  governed  by the  6-order  differential operator that factorizes, as it is easy to see,  into the product of three
  2nd order Lichnerowitz-type   operators  if the background is an Einstein one.
  Restricted to transverse traceless $h_{\mu\nu}$  the kinetic operator in \rf{B.2} 
   is (cf. \rf{423}) 
  $- D^6 + ...= \Delta_{2\,\perp} ( 8  )\ \Delta_{2\,\perp} ( 6  ) \ \Delta_{2\,\perp} ( 2  )$   with 
  \be 
  \Delta_{2} (M^2)\, h_{\m\nu}  = \big( -D^{2}  + \tfrac{1}{30} M^2 { R } \big)\,h_{\mu\nu}-2\,C_{\mu\rho\nu\sigma}\,h^{\rho\sigma}  \ ,
  \ee 
  where $R$ is the scalar   curvature    and $C$ is the Weyl tensor. 
  On  a   unit-radius $S^6$   where $R=30$ and $ \ C=0$ we  then get the one-loop partition function in \rf{423}.

    This factorization implies that   the one-loop 
 conformal anomaly coefficients  in \rf{4.1}  corresponding  to \rf{B.2} 
 can be computed     following  \ci{Bastianelli:2000hi} by 
using directly  the  general expression \ci{Gilkey:1975iq} 
for the $b_6$   Seeley-DeWitt  coefficient of  the corresponding    2nd order Laplace-type 
operator $\Delta= - D^2   + X $ defined on $k\leq 2$ tensors
  that enter the  generalization of the $S^6$ partition function in \rf{423}.

To simplify the computation one  may use a shortcut  and  consider several  special backgrounds. 
Considering $S^6$    case one can easily determine the value of a-coefficient. 
For   a  symmetric-space  Einstein  background  with a non-zero Weyl tensor 
(where $I_i$ invariants satisfy one linear relation) 
 one  is able  to  find   $\cc_1$ and $\cc_3$ in terms of $\cc_2$   \cite{Pang:2012rd}:
\be
\la{B.3}
\text{a} = \tfrac{3005}{2\times 7!}, \qquad\qquad  \  \text{c}_{1} = \tfrac{5633}{105}-4\,\text{c}_{2}, \qquad\ \  \ \ \ 
\text{c}_{3} = -\tfrac{35543}{5040}+\tfrac{5}{8}\,\text{c}_{2}.
\ee
The value of  $\text{c}_{2}$ may be fixed by  considering   the case of a 
 Ricci flat   background  where  the  partition function takes a simple form 
\be
\la{B.4}
Z_{2} = \Big[\frac{(\det\Delta_{1})^{4}}{(\det\Delta_{2})^{3}}\Big]^{1/2}\ ,
\ee
with  $\Delta_1,\Delta_2$   being the   standard   Laplacians acting 
on unconstrained  vector and traceless  tensor with $R_{\mu\nu}=0$, \foot{To compare, the usual Einstein  theory partition function on a Ricci flat background is 

\noindent
$
Z_{2\, E} = \big[\frac{\det\Delta_{1}}{\det\Delta_{2}\  \det \Delta_0}\big]^{1/2}\ 
$.} 
$\Delta_1 h_\mu = - D^2 h_\mu \ , \ \ 
\Delta_{2}\,h_{\mu\nu} = -D^{2}\,h_{\mu\nu}-2\,C_{\mu\rho\nu\sigma}\,h^{\rho\sigma}
$.
To find the corresponding   $b_6$  coefficient  
 for the vector Laplacian  $\Delta_1$  from the general  expressions in 
  \ci{Gilkey:1975iq,Bastianelli:2000hi}
 one is to use  that here  
the  covariant   derivative contains an  extra "internal"  vector connection part with the 
 curvature $(\F_{\mu\nu })_{\a}^{\ \ \b}=C\indices{_{\mu\nu\a}^{\b}}$. Then one finds 
  \begin{align}
\la{B.5}
 7!\,\bar b_{6} \big[ \D_1\big] 
 = & \tfrac{80}{9} C_{\a}{}^{\m}{}_{\g}{}^{\n} C^{\a\b\g\d} \, C_{\b\m\d\n}   
 -  \tfrac{164}{3} C_{\a\b}{}^{\m\n} C^{\a\b\g\d} \, C_{\g\d\m\n} \notag\\ &
 -96 C^{\a\b\g\d} D^{2}C_{\a\b\g\d} 
 -58 (D_{\m}C_{\a\b\g\d})^{2}.
\end{align}
In the case of spin 2   operator $-D^2 + X$  one has  $(\mc F_{\mu\nu})_{\alpha\beta, \rho\sigma} = \tfrac{1}{2}
C_{\mu\nu\alpha\rho}g_{\beta\sigma}+\dots
$
(dots  stand for  3 similar   terms that symmetrize in $(\alpha\beta)$ and $(\rho\sigma)$)
and $ X_{\mu\nu, \rho\sigma} =-  C_{\mu\rho\nu\sigma}- C_{\mu\sigma\nu\rho}.$
A straightforward computation  then gives 
 \begin{align}
\la{B.9}
 7!\,b_{6} \big[ \D_2\big] 
 =  &\tfrac{49984}{9} C_{\a}{}^{\m}{}_{\g}{}^{\n} C^{\a\b\g\d} \, C_{\b\m\d\n}   
 -  \tfrac{1388}{9} C_{\a\b}{}^{\m\n} C^{\a\b\g\d} \, C_{\g\d\m\n} \notag\\
 &+1416\, C^{\a\b\g\d} D^{2}C_{\a\b\g\d} 
 +544\, (D_{\m}C_{\a\b\g\d})^{2}.
 \end{align}
The full Seeley-DeWitt   coefficient  corresponding to \rf{B.4}  is  given  by 
 the combination $3\,b_{6}[\D_{2}]-4\,b_{6}[\D_{1}]$  and thus 
contains   one of the total derivative terms  discussed in  \cite{Bastianelli:2000hi}.
Ignoring total derivative terms  and using 
 the relations  between the  invariants in \rf{4.1}  that exist   in  the Ricci flat   case 
 ($E_6 = 32 (2 I_1 + I_2) , \ I_3 = 4 I_1 - I_2$) 
 one has  in   general   
\be
\la{B.10}
\bar b_{6} \Big|_{R_{\mu\nu}=0} = -\big[\text{a}-\tfrac{1}{192}\,(\text{c}_{1}+4\,\text{c}_{2})\big]\,E_{6}+
(\text{c}_{1}-2\text{c}_{2}+6\text{c}_{3})\,I_{1} \ . 
\ee
As a result, we find 
\be
\la{B.11}
 \text{a}-\tfrac{1}{192}\,(\text{c}_{1}+4\,\text{c}_{2}) = \tfrac{377}{20160}, \qquad\qquad 
 \text{c}_{1}-2\text{c}_{2}+6\text{c}_{3} = -\tfrac{1}{210}.
 \ee
  Combining \rf{B.11}  with \rf{B.3}    we conclude  that the first   relation is   satisfied identically 
  while   the second one  determines $\cc_2$. Thus  
   finally 
 \be
 \text{c}_{1} = \tfrac{1507}{45},\qquad\qquad  \text{c}_{2} = \tfrac{635}{126}, \qquad\qquad  \text{c}_{3} = 
 -\tfrac{1639}{420}.\la{A11}
 \ee


\section{Massless  higher spins  on $S^{4}_{q}$ \la{JJ}}

One  may define free  massless  higher spin (MHS)   fields  on   $S^4$  and consider, as in the case of AdS$_4$ \ci{Giombi:2013fka}, 
the corresponding partition function
  built  out of the  determinants of the 
operators $\Delta_{s\, \perp} (M^2) $ 
 on spin $s$ TT tensors  \rf{21}\foot{The analytic continuation from AdS$_4$ to $S^4$ corresponds to 
changing the sign of the   square of the curvature radius or $M^2 \to - M^2$.  Note that  the  $s=0$ case is  a  special  case of $Z_s$ assuming  one drops   the  ghost   contribution. Here we set the radius of $S^4$ to 1.}
\be    \la{2222} 
Z_{0}=  \Big[\frac{1}{\det\Delta_{0}(2)}\Big]^{1/2} 
 \ , \qquad \qquad  Z_{s}= \Big[\frac{\det\Delta_{s-1\, \perp}(1-s^2)}{\det\Delta_{s\, \perp}(2+ 2s - s^2)} 
\Big]^{1/2}  \ . \ee
Then    extending  these  MHS partition functions 
to  $S^4_q$  we    may  use the results in \rf{288} and Appendix \ref{app:den}    to  compute 
the corresponding total  $\wz(q)$ function which is the coefficient of the  log UV divergence. 
 The expressions for the scalar   and spin 1 field are  the same as in \rf{200}   and \rf{201}, while for $s=2,3,4$ we obtain (cf. \rf{201}--\rf{2.15}) 
\begin{align}
\la{J1}
\wz_2(q)= &
-\frac{1}{180\,q^{3}}-\frac{2}{q^{2}}+\frac{10}{3\,q}-\frac{22}{3}-\frac{401}{60}\,q \ , \notag\\
\wz_3(q) =  &  
-\frac{1}{180\,q^{3}}-\frac{12}{q^{2}}+\frac{45}{2\,q}-39
-\frac{2251}{60}\,q \ ,  \\
\wz_4(q) =&  
-\frac{1}{180\,q^{3}}-\frac{40}{q^{2}}+\frac{232}{3\,q} -\frac{376}{3}
-\frac{7361}{60}\,q \ .  \notag \end{align}
Here we used   \rf{344} 
with the number of zero  modes   being (cf. \rf{333})
\be
\la{J3}
n_{\text{z}, s} = \text{k}_{s}-\text{k}_{s-1} = \tfrac{1}{6}\,s^{2}\,(1+5\,s^{2}) \ . 
\ee
A natural generalization  to any $s >0$  is then (cf. \rf{3115})
\be
\la{J2}
\widehat\zeta_{s}(q) = -\frac{1}{180\,q^{3}}+\frac{s^{2}(1-s^{2})}{6\,q^{2}}
-\frac{s^{2}(3-2s^{2})}{6\,q}+\frac{s^{2}(1-3s^{2})}{6}-\frac{1}{60}(1-20\,s^{2}+30\,s^{4})\,q.
\ee
This   has the expected   general structure \rf{2.21}   with the number of degrees of freedom $\nu=2$ 
and 
$E_{c,s}$   being  the Casimir energy of the MHS field on $\mathbb R \times S^{3}$
(see  eq.(5.7) in \cite{Giombi:2014yra}).\footnote{One can obtain $E_{c,s}$ from the single particle partition function $\mc Z_{s}(x) = \frac{x^{s+1}}{(1-x)^{3}}\,\big[2s+1-(2s-1)\,x\big]$
where  $x= e^{-\beta}$   using eq. (5.16)  in \cite{Giombi:2014yra}.}

The coefficient of the total UV   log divergence of the tower of massless higher spin fields 
on $S^4$ is   given by 
$\wz_0(1) + \sum_{s=1}^\infty \wz_s(1)$  and vanishes when regularized 
with an exponential cutoff   or zeta-function    \cite{Giombi:2013fka,Giombi:2016ejx}.
The same is true for the sum of the Casimir energies \cite{Giombi:2014yra}.


\iffa
\be
\la{J4}
\text{a}_{s} = \tfrac{1}{360}\,(2-15\,s^{2}+75\,s^{4}),
\ee
in agreement with Eq.(227) of \cite{Giombi:2016ejx}. 
Similarly, we can also evaluate the following 
``c-anomaly''
\be
\la{J5}
\left. -\tfrac{1}{4}[q\,\widehat\zeta^{\text{MHS}}(q)]''\right|_{q=1} = \tfrac{1}{60}\,(1-15\,s^{2}+20\,s^{4}).
\ee 
Notice also that 
\fi 
Similarly, on   the  conical  $S^4_q$   space   we find  that  the
  total $\wz(q)$  function  also   vanishes, i.e. the regularized sum\foot{Note that for massless  higher spins in $d$ dimensions  the regularization prescription  is with  cutoff factor 
 $e^{-\eps ( s + {d-4\ov 2})} $  \ci{Giombi:2014iua}. For conformal  higher spins 
 one has instead $e^{-\eps ( s + {d-3\ov 2})} $  as   they are effectively associated with the boundary, i.e. 
 one is to replace $d \to d-1$.}
\begin{align}
\la{J6}
& \widehat\zeta_{0}(q) + \sum_{s=1}^{\infty}e^{-\eps\,s}\, \widehat\zeta_{s}(q) 
\notag \\
&\qquad \qquad = \frac{-\frac{4}{q^2}-12 q+\frac{8}{q}+8}{\eps ^5}+\frac{\frac{1}{3 q^2}+\frac{2
   q}{3}-\frac{1}{q}+\frac{2}{3}}{\eps ^3}+\frac{-\frac{1}{180 q^3}-\frac{q}{60}}{\eps }+ 0 +  {\cal O}(\eps),
\end{align}
has zero  finite part.
Then  the  sum of  free energies on $S^1_q \times \HH^3$  \rf{2.19}  and  thus  the   \renyi   entropies  \rf{36}
also vanish. Such formally   defined 
\renyi  entropy   may be associated to  the  tower of massless higher spins
in flat space   and thus its   vanishing   is   consistent  with  "topological" nature  of such higher spin  theory
 \cite{Beccaria:2015vaa}.


\section{$B_{2}$ Seeley-DeWitt   coefficient for 4d spins  $s\leq 3$ \la{BBB} }

It is of interest to compare  consequences of our expressions for the zeta-functions in section 2 
 with some  previous results in \ci{Fursaev:1996uz}.
Given  a spectral  zeta-function $\zeta(z; q)$  for the operators 
$\Delta_{s\, \perp}(M^2)$  we can also extract the $B_2$ 
Seeley-DeWitt  coefficient (of quadratic UV divergences)   that appears 
 in the $t\to 0$   expansion of the heat kernel in 4 dimensions, 
$K(t) = B_0 t^{-2} + B_2 t^{-1} + B_4 + O(t)$.  After  a convenient rescaling   we  have   
 \be\la{ff1}
 \widetilde B_{2} \equiv  \frac{(4\pi)^{2}}{\text{Vol}(S^{4})}\,B_{2} = 6 \, B_2
 = 6 \lim_{z\to 1}\, (z-1)\,\zeta(z; q) \ .
 \ee
 Reintroducing the factors of the scalar curvature (equal to  12 for  a unit-radius $S^{4}$)
we   have   from \rf{288}
 \begin{align}
 \la{Q.3}
& \widetilde B_{2}^{(0)} (q, M^{2})\te = \frac{R}{24\,q}+q\,\left(\frac{R}{8}-M^{2}\right), \qquad \quad
 \widetilde B_{2}^{(1\,\perp)} (q, M^{2}) 
 \te = -\frac{R}{2}+\frac{R}{8\,q}+q\,
 \left(\frac{5R}{8}-3M^{2}\right),\ \notag\\
 &\widetilde B_{2}^{(2\,\perp)}(q, M^{2})  \te = -\frac{5R}{2}+\frac{5R}{24\,q}+q\,\left(
 \frac{35R}{24}-5M^{2}\right) ,\\
&  \widetilde B_{2}^{(3\,\perp)}(q, M^{2})  \te = -7 R+\frac{7R}{24\,q}+q\,\left(
 \frac{21R}{8}-7M^{2}\right) \ . \quad\notag
 \end{align}
 Using the relations in Appendix  A of \cite{Tseytlin:2013jya},
 we then  find  for the   coefficients in heat kernels 
 of operators $\Delta_{s} (M^2) $  defined  on fields  without the transversality condition 
 \begin{align}
 \widetilde B_{2}^{(1)}(q, M^{2}) &= \te  \widetilde B_{2}^{(1\, \perp)}(q, M^{2})
 + \widetilde B_{2}^{(0)}(q, M^{2}-3)= -\frac{R}{2}+\frac{R}{6\,q}+q\,\left(
 -4\,M^{2}+R\right)
 , \la{f3} \\
 \widetilde B_{2}^{(2)}(q, M^{2}) &=  \widetilde B_{2}^{(2\, \perp)}(q, M^{2})
 + \widetilde B_{2}^{(1\, \perp)}(q, M^{2}-5)+ \widetilde B_{2}^{(0)}(q, M^{2}-8) \notag\\
 &=  \te  -3\,R+\frac{3R}{8\,q}+q\,\left(
 -9\,M^{2}+\frac{33R}{8}\right),\la{f4}
 \\
 \widetilde B_{2}^{(3)}(q, M^{2}) &=  \widetilde B_{2}^{(3\, \perp)}(q, M^{2})
 + \widetilde B_{2}^{(2\, \perp)}(q, M^{2}-7)+ \widetilde B_{2}^{(1\perp)}(q, M^{2}-12)
 +\widetilde B_{2}^{(0)}(q, M^{2}-15)\notag\\
 &\te =  -10\,R+\frac{2R}{3\,q}+q\,\left(
 -16\,M^{2}+12 R\right).\la{f5}
  \end{align}
It is convenient to split the coefficient $\widetilde B_{2}$ into a regular  "bulk"  part and  "surface" part coming from the conical singularity (cf.  also discussion in Introduction) 
\be
\la{Q.6}
\widetilde B_{2}(q, M^{2}) = \underbrace{q\,\widetilde B_{2}(1, M^{2})}_{\rm "bulk"}+
\underbrace{ \big[
\widetilde B_{2}(q, M^{2}) -q\,\widetilde B_{2}(1, M^{2}) \big]}_{\rm "surface"}.
\ee
The "bulk" part is  given by the usual Seeley-DeWitt coefficient evaluated on $S^{4}_{q}$ with the
singular region excised:
  it is   given  by the  standard  $S^4$ (i.e. $q=1$) expression
  \be \te \widetilde B_{2}(1, M^{2}) = N_{s}\,\big(\frac{R}{6}-M^{2}\big), \   \qquad \qquad 
  N_s=(s+1)^2\la{Q.7} \ , \ee
   times the $q$-factor
 which accounts for the  
volume of $S^{4}_{q}$.
The "surface" part in (\ref{Q.6}) vanishes  for $q=1$ by construction.\footnote{In the
approaches that represent the conical singularity  in terms of a singular part in the 
 curvature 
the "surface" term   originates  from  an integral over $S^{2}$
as near the cone singularity $S^{4}_{q}\sim
\mathscr C_{q}\times S^{2}$.}
The splitting (\ref{Q.6}) then takes the form 
\begin{align}
\la{Q.8}
\widetilde B_{2}^{(0)}(q, M^{2})  &\te = q\,\big(\frac{R}{6}-M^{2}\big)+\frac{R}{24}\big(
\frac{1}{q}-q\big),\notag\\
\widetilde B_{2}^{(1)}(q, M^{2})  &\te = q\,\big(\frac{2R}{3}-4M^{2}\big)+\big(
-\frac{R}{2}+\frac{R}{6\,q}+\frac{R}{3}\,q\big), \notag \\
\widetilde B_{2}^{(2)}(q, M^{2})  &\te = q\,\big(\frac{3R}{2}-9 M^{2}\big)+\big(
-3R+\frac{3R}{8\,q}+\frac{21 R}{8}\,q\big), \notag \\
\widetilde B_{2}^{(3)}(q, M^{2})  &\te = q\,\big(\frac{8R}{3}-16 M^{2}\big)+\big(
-10 R+\frac{2 R}{3\,q}+\frac{28 R}{3}\,q\big).
\end{align}
The  spin 0,1,2   "surface" terms in \rf{Q.8}  may  be compared with the
results of \cite{Fursaev:1996uz} (see   eqs.(2.8), (2.11), (2.13) there  with $\beta=2\pi q$)\footnote{We
 factor out the  volume ratio $\frac{\text{Vol}(S^{2})}{\text{Vol}(S^{4})}$ that takes into
account that the surface term in \cite{Fursaev:1996uz} is integrated over $S^{2}$.
Notice also that the  expression in eq.  (2.13) of \cite{Fursaev:1996uz} has an additional $8\pi$ term
that remains even in
the smooth $q\to 1$ limit. In the above comparison we did not include this contribution.
It is due to the dipole modes discussed in that paper. These are
normalizable modes that exist for $q<1$. They have a wave-function which is summable but
singular at  the cone's apex. We do not see these 
 modes because we constructed the spectral $\zeta$ function
by considering  as   boundary conditions  that 
the  eigentensors of the Laplace-type  operator  are
regular everywhere,  {\em i.e.} the analogue of the Friedrich extension, see for instance  
  section 1.5 of \cite{Iellici:1998ce}.}
\begin{align}
\la{Q.9}
s=0:\ &\te  \frac{\beta}{6}\big[ (\frac{2\pi}{\beta})^{2}-1\big]\, 
\frac{\text{Vol}(S^{2})}{\text{Vol}(S^{4})}  = \frac{1}{2q}-\frac{q}{2},\notag \\
s=1:\ &\te  \left[N_1\  \frac{\beta}{6}\,  \big[ (\frac{2\pi}{\beta})^{2}-1\big]\, 
+2\,(\beta-2\pi)\right]
\frac{\text{Vol}(S^{2})}{\text{Vol}(S^{4})} =
-6+\frac{2}{q}+4q,\notag \\
s=2: \ &\te  \left[N_2\ \frac{\beta}{6}\,  \big[ (\frac{2\pi}{\beta})^{2}-1\big]\,  
+12\,(\beta-2\pi)\right]
\frac{\text{Vol}(S^{2})}{\text{Vol}(S^{4})} =
-36+\frac{9}{2q}+\frac{63}{2}q\ . 
\end{align}
These   match the  "surface" terms in \rf{Q.8} after setting $R=12$. 
The  extension of the pattern in  the l.h.s. of \rf{Q.9} to the spin 3  case 
 that matches  the $s=3$  expression in \rf{Q.8}  is 
\be
s=3: \  \te \left[N_3\,  \frac{\beta}{6}\,  \big[ (\frac{2\pi}{\beta})^{2}-1\big]\   +40\,(\beta-2\pi)\right]
\frac{\text{Vol}(S^{2})}{\text{Vol}(S^{4})} =
-120+\frac{8}{q}+112\,q.
\ee

\bibliography{BT-Biblio}
\bibliographystyle{JHEP}

\end{document}